\documentclass[10pt, onecolumn, conference]{IEEEtran}

\usepackage{graphicx}
\usepackage{booktabs}

\usepackage[misc,geometry]{ifsym} 
\usepackage{academicons}
\usepackage{color}
\usepackage{hyperref} 
\usepackage{caption}
\usepackage{amsmath}
\usepackage{aas_macros}
\usepackage[bottom]{footmisc}
\usepackage{supertabular}
\usepackage{afterpage}
\usepackage{url}
\usepackage{pifont}
\usepackage{multicol}
\usepackage{multirow}
\usepackage{xspace}
\usepackage{float}
\usepackage{footmisc}

\usepackage{appendix}

\definecolor{orcidlogo}{rgb}{0.37,0.48,0.13}
\definecolor{unilogo}{rgb}{0.16, 0.26, 0.58}
\definecolor{maillogo}{rgb}{0.58, 0.16, 0.26}
\definecolor{darkblue}{rgb}{0.0,0.0,0.0}
\hypersetup{colorlinks,breaklinks,
            linkcolor=darkblue,urlcolor=darkblue,
            anchorcolor=darkblue,citecolor=darkblue}
            
\DeclareCaptionType{equ}[][]

\title{Synthetic Data: AI's New Weapon Against Android Malware}

\author{
\IEEEauthorblockN{
    Angelo Gaspar Diniz Nogueira\IEEEauthorrefmark{1},
    Kayua Oleques Paim\IEEEauthorrefmark{2},
    Hendrio Bragança\IEEEauthorrefmark{3} \\
    Rodrigo Brandão Mansilha\IEEEauthorrefmark{1},
    Diego Kreutz\IEEEauthorrefmark{1}
}
\IEEEauthorblockA{\IEEEauthorrefmark{1}Federal University of Pampa (UNIPAMPA)}
\IEEEauthorblockA{\IEEEauthorrefmark{2}Federal University of Rio Grande do Sul (UFRGS)}
\IEEEauthorblockA{\IEEEauthorrefmark{3}Federal University of Amazonas (UFAM)}
}

\newcommand{\toolname}{MalSynGen\xspace}

\begin{document}

\maketitle


\begin{abstract}
The ever-increasing number of Android devices and the accelerated evolution of malware, reaching over 35 million samples by 2024, highlight the critical importance of effective detection methods. Attackers are now using Artificial Intelligence to create sophisticated malware variations that can easily evade traditional detection techniques. Although machine learning has shown promise in malware classification, its success relies heavily on the availability of up-to-date, high-quality datasets. The scarcity and high cost of obtaining and labeling real malware samples presents significant challenges in developing robust detection models. In this paper, we propose MalSynGen, a Malware Synthetic Data Generation methodology that uses a conditional Generative Adversarial Network (cGAN) to generate synthetic tabular data. This data preserves the statistical properties of real-world data and improves the performance of Android malware classifiers. We evaluated the effectiveness of this approach using various datasets and metrics that assess the fidelity of the generated data, its utility in classification, and the computational efficiency of the process. Our experiments demonstrate that MalSynGen can generalize across different datasets, providing a viable solution to address the issues of obsolescence and low quality data in malware detection.
\end{abstract}



\section{Introduction}
\label{sec:intro}

With approximately 3 billion Android devices in operation worldwide \cite{theverge_android_2021}, the mobile cybersecurity landscape faces formidable challenges. In 2024 alone, Kaspersky reported over 33.3 million cyberattacks targeting smartphone users globally, encompassing diverse forms of malware and unwanted software \cite{kaspersky_mobile_threats_2024}.
\color{black}
Adding to this problem, attackers are using Artificial Intelligence (AI) to rapidly generate new malware variants by exploiting patterns learned from existing malware \cite{renjith2022gang}. Furthermore, AI generators can potentially learn from benign applications, enabling malware to evade advanced antivirus solutions. Consequently, there is a pressing need for innovative techniques that can not only detect current threats, but also anticipate future ones.

Recent research demonstrates the effectiveness of supervised learning techniques in detecting and mitigating threats to Android malware, achieving high detection rates when combined with robust feature engineering \cite{kouliaridis2021comprehensive, meijin2022systematic}. However, the performance of these methods depends on: (1) the quantity of training samples, ensuring a sufficiently representative dataset; (2) the representativeness of attributes, encompassing the diversity and volume of features; and (3) the currency of the dataset, reflecting the latest trends and technological advances of malware \cite{botacin2021challenges, paullada2021data}.

The scarcity of high-quality training data is a primary factor that contributes to the failure of 80\% of AI projects \cite{ai2023top}. Specifically, sample size, feature representativeness, and data recency directly influence the performance of Android malware detection models \cite{miranda2022debiasing,kouliaridis2020feature,wang2019rmvdroid}. This challenge is exacerbated by the increasing use of generative AI technologies, such as Generative Adversarial Networks (GANs), by attackers to rapidly and continuously mutate malware \cite{xiao2018generating,chauhan2021polymorphic,hu2022generating}.

Obtaining sufficient, representative, and up-to-date real Android malware samples is a complex, costly, and time-consuming undertaking \cite{vilanova2022adbuilder, siqueira2021quickautoml, siqueira2022avaliaccao, assolin2022droidautoml, Rocha2023AMGenerator}. Limitations such as the difficulty of acquiring malicious samples and the labor intensive labeling process, exemplified by the restricted labeling rate of services such as VirusTotal\footnote{VirusTotal, a leading API service for metadata labeling, limits processing to 250 labels per day: \url{https://developers.virustotal.com/reference/overview}.}, render the continuous creation of updated datasets for model training and validation technically impractical.


Synthetic data generation has emerged as a valuable solution to address data scarcity across various domains. While numerous generic tools exist for generating synthetic tabular data \cite{xu2019modeling,rajabi2022tabfairgan}, specific applications within the Android malware domain remain unexplored. This domain presents two key challenges: hyperparameter selection and synthetic data evaluation.
Firstly, the effectiveness of machine learning algorithms, particularly deep learning models like cGANs, is highly sensitive to hyperparameter tuning \cite{weerts2020importance,kurach2019large,sabiri2022effect}. Secondly, there is a lack of standardized metrics for evaluating synthetic data, both generally \cite{HoldoutFrontiers2021} and within the specific context of Android malware.
Ideally, synthetic data should maintain statistical fidelity to the original data while enhancing its utility for training supervised learning models. Furthermore, the generation process must be computationally efficient, offering a cost-effective alternative to acquiring real data.

This paper introduces MalSynGen (Malware Synthetic Data Generation), a comprehensive methodology and publicly available framework for generating and evaluating synthetic tabular data tailored for Android malware detection. We utilize a conditional Generative Adversarial Network (cGAN) model, inspired by \cite{mirza2014conditional}, to generate synthetic data. Our evaluation framework assesses: (i) the \textit{fidelity} of the generated synthetic data compared to real data, and (ii) the \textit{utility} of the synthetic data in Android malware classification using a variety of classifiers. This work significantly expands upon our prior research by providing a more detailed methodology, a broader evaluation across diverse Android malware datasets, and a thorough analysis of computational resource consumption.

The key contributions of this expanded research are:
\begin{enumerate}
    \item A cGAN model designed to generate synthetic tabular data that effectively supports Android malware classification.
    \item A systematic methodology for training and evaluating the proposed cGAN model.
    \item A comprehensive set of metrics for assessing both the \textit{fidelity} and utility of the generated synthetic data in Android malware classification.
    \item An extensive evaluation across multiple established Android malware datasets, demonstrating the generalization capabilities of our methodology.
\end{enumerate}

The remainder of this paper is structured as follows: Section \ref{sec:relatedwork} reviews related work. Section \ref{sec:conceitual} details the conceptual methodology and cGAN model. Section \ref{sec:evaluation} outlines the evaluation process, including implementation and deployment details. Section \ref{sec:results} presents and discusses the evaluation results. Finally, Section \ref{sec:conclusion} summarizes the main conclusions and outlines future research directions.

\color{black}


\section{Related Work}\label{sec:relatedwork}

In Table \ref{tab_trabalhos_relacionados}, we present the main related works in the context of tabular data augmentation. We list the techniques, metrics, domain, and datasets used in each work. As it can be seen, GANs are frequently used to generate synthetic tabular data.
However, other techniques are also used in the data augmentation process, such as the use of large language models (LLMs) \cite{borisov2022language} and diffusion models \cite{villaizan2024diffusion}, due to their ability to effectively capture and generate complex patterns.

\begin{table*}[!ht]
\small
\centering
\caption{Related works on the context of synthetic tabular data. We list the Technique, metrics domain and datasets}
\label{tab_trabalhos_relacionados}
 \resizebox{\textwidth}{!}{%
\begin{tabular}{ccccc}
\hline
\textbf{Publication} &
\textbf{Technique} &
\textbf{Metrics} &
\begin{tabular}[c]{@{}c@{}} \textbf{Evaluation}\\ \textbf{Method}\end{tabular} &
\textbf{Datasets} \\ \hline
\cite{choi2017generating} &
GAN &
\begin{tabular}[c]{@{}c@{}}Precision, F1 score, recall,\\ Bernoulli distribution\end{tabular} &
TSTR &
2 medical \\ \hline
\cite{xu2018synthesizing} &
GAN &
\begin{tabular}[c]{@{}c@{}}F1 score, accuracy, \\ mean squared error, and absolute squared error\end{tabular} &
TSTR &
3 general purpose \\ \hline
\cite{park2018data} &
GAN &
\begin{tabular}[c]{@{}c@{}}Cumulative distribution,\\ F1 score, mean relative error,\\ Euclidean distance, and AUC\end{tabular} &
TSTR &
4 general purpose \\ \hline
\cite{xu2019modeling} &
cGAN &
\begin{tabular}[c]{@{}c@{}}Accuracy, F1 score,\\ R², and log-likelihood\end{tabular} &
TSTR &
7 general purpose \\ \hline
\cite{mimura2020using} &
GAN &
\begin{tabular}[c]{@{}c@{}}Accuracy,\\ F1 score, and Recall\end{tabular} &
TSTR &
1 VBA malware \\ \hline
\cite{rajabi2022tabfairgan} &
GAN &
\begin{tabular}[c]{@{}c@{}}Accuracy, \\ F1 score, and measure of discrimination\end{tabular} &
TSTR &
5 demographic data \\ \hline
\cite{amin2022Android} &
GAN &
\begin{tabular}[c]{@{}c@{}}F1 score, recall, accuracy,\\ precision, AUC, FPR, and coverage\end{tabular} &
TSTR &
2 Android malware \\ \hline
\cite{borisov2022language} &
LLM &
\begin{tabular}[c]{@{}c@{}}Machine learning efficiency,\\ accuracy, distance to the \\ closest registries, measure of discrimination, \\ and\\ bivariate joint distribution chart\end{tabular} &
TSTR &
6 general purpose \\ \hline
\cite{Casola2023DroidAugmentor} &
cGAN &
\begin{tabular}[c]{@{}c@{}}F1 score, recall, \\ accuracy, precision,\\ KL divergence, maximum \\ mean discrepancy, and mean squared error\end{tabular} &
TSTR &
1 Android malware \\ \hline
\cite{li2024syndroid} &
cGAN &
\begin{tabular}[c]{@{}c@{}}F1 score, recall, \\ accuracy, and precision\end{tabular} &
TSTR &
2 Android malware \\ \hline
\cite{zhao2024ctab} &
cGAN &
\begin{tabular}[c]{@{}c@{}}Accuracy, F1 score, AUC,\\ mean absolute error,\\ R² score, and variance score\end{tabular} &
TSTR &
6 general purpose \\ \hline
\cite{villaizan2024diffusion} &
\begin{tabular}[c]{@{}c@{}}Diffusion\\ Model\end{tabular} &
\begin{tabular}[c]{@{}c@{}}Average Wasserstein distance, \\ Mean Jensen-Shannon distance,\\ Mean L2 distance \\ of the correlation matrices, \\ Machine learning efficiency, \\ and F1 score\end{tabular} &
TSTR &
7 general purpose \\ \hline
\textbf{This work} &
cGAN &
\begin{tabular}[c]{@{}c@{}}F1 score, recall, accuracy,\\ precision, AUC, \\ cosine similarity, squared error, \\ $p$-value, mean discrepancy, and \\ Euclidean distance\end{tabular} &
TSTR/ TRTS &
7 Android malware \\ \hline
\end{tabular}
}
\end{table*}

Most solutions for tabular data seek to capture the particularities of a specific context, such as healthcare \cite{choi2017generating}, demographics \cite{rajabi2022tabfairgan} and malware VBA \cite{mimura2020using}.
We can also observe that three solutions are specific to the context of malware Android, where we can see a predominance of cGANs \cite{amin2022Android, li2024syndroid}.

While prior solutions largely evaluate synthetic data utility through supervised learning model performance, they predominantly rely on standard binary classification metrics like precision, accuracy, recall, and F1-score. Furthermore, they typically utilize synthetic data in either the training or evaluation phase, but not both. This approach presents challenges, as high performance could stem from mere data replication rather than genuine novelty, while entirely novel data might yield poor classification if lacking inherent structure.

To overcome these limitations, we expand upon existing metrics by proposing two distinct categories: \textit{utility} and \textit{fidelity}. Utility metrics align with those commonly used in related works, whereas \textit{fidelity} metrics, as emphasized in recent research \cite{canbek2021benchmetrics,rainio2024evaluation}, are specifically designed for assessing generative models and synthetic data quality. Additionally, we integrate synthetic data into complementary evaluation methodologies to enhance robustness. Unlike previous works that primarily use the Training on Synthetic, Testing on Real (TSTR) method, we adopt a dual approach: Training on Synthetic, Testing on Real (TSTR) and Training on Real, Testing on Synthetic (TRTS). This comprehensive strategy ensures a more thorough evaluation.

\section{Malware Synthetic Data Generation (MalSynGen)}
\label{sec:conceitual}

This section details the conceptual components of the MalSynGen framework, encompassing both the overall methodology and the underlying cGAN model. We begin by outlining the methodological process and subsequently describe the generative model.

\subsection{Methodology}

\begin{figure*}[!htp]
    \centering
    \includegraphics[width=1.0\textwidth]{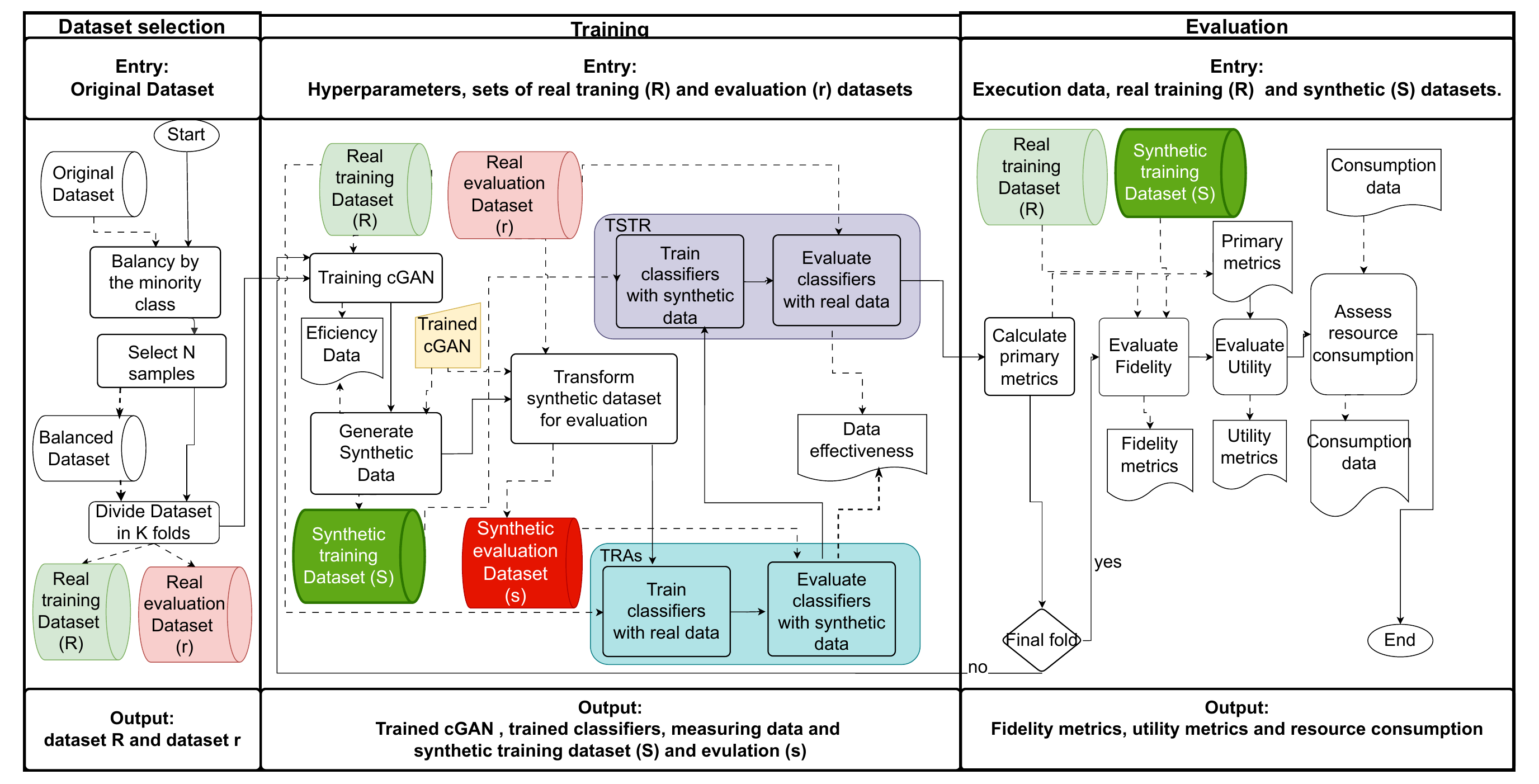}
    \caption{Selection, training, and evaluation of cGANs using MalSynGen. Solid arrows denote the sequential order of processes, while dotted arrows indicate the flow of artifact creation or utilization (datasets, models, raw data, and performance metrics). Certain artifacts are depicted multiple times to streamline the visualization of arrow intersections.}
    \label{fig_fluxo_de_execucao}
\end{figure*}

We illustrate the execution flow of the \toolname framework in Figure~\ref{fig_fluxo_de_execucao}. The proposed flow consists of three main steps: selection and manipulation of the original dataset, training of classifiers, training of the conditional Generative Adversarial Network (cGAN), and evaluation of results.

In the first stage, \textbf{selection}, we choose a real dataset and perform balancing by the class (benign or malignant) with the fewest samples  between the two categories.
The balancing of the benign and malignant samples of the dataset is accomplished through the use of subsampling techniques.
We then prepare the \emph{dataset} for k-folds cross-validation.
The balanced real \emph{dataset} is divided into \textit{k} equally sized subsets, and at each iteration, one part is chosen as the evaluation subset (\emph{Dataset} \textsf{r}) and the remaining \textit{(k-1)} subsets are used for training (\emph{Dataset} \textsf{R}).

In the \textbf{training} step, the framework receives as input the cGAN training hyperparameters\footnote{In this work, we implement our own cGAN model.} and the hyperparameters of the classification algorithms. In addition, for each fold, the respective training (\textsf{R}) and evaluation (\textsf{r}) subsets are used.
The training dataset (\textsf{R}) and part of the hyperparameters are used to train the cGAN generative neural network. The trained cGAN instance is then used to generate a synthetic training dataset (\textsf{S}). In sequence, the same cGAN instance is used to generate a synthetic evaluation dataset (\textsf{s}) by performing a transformation over the real evaluation subset dataset (\textsf{r}). It is important to note that the evaluation data and synthetic evaluation data are not used at any point to train or fine-tune the cGAN generative model.

In addition to training and using cGANs to generate synthetic data, we also need to select, tune, and train classifiers (e.g., Support Vector Machine, Decision Tree) to subsequently assess the utility of the generated synthetic data. We implement the two methods proposed in \cite{esteban2017real}: training on real data (\textsf{R}) and test the classifier with synthetic data (\textsf{s}) (TRTS) and; training on synthetic data (\textsf{S}) and test with real data (\textsf{r}) (TSTR).

In the \textbf{evaluation} stage, we execute the classifiers and compute the primary metrics, \textit{utility} and \textit{fidelity}.

For each trained classifier and corresponding evaluation subset, we extract the binary classification test results, specifically the counts of true positives, false positives, true negatives, and false negatives.

From these binary classification results, we construct confusion matrices and calculate standard binary classification metrics, including Accuracy, Precision, Recall, AUC, and F1-Score. Given that false negatives are more detrimental than false positives in malware detection, we prioritize Recall over Precision.

Upon completing the k-fold cross-validation, we proceed to the \textit{utility} evaluation. This involves calculating the mean and standard deviation of each primary metric for each classifier, using both the Training on Synthetic, Testing on Real (TSTR) and Training on Real, Testing on Synthetic (TRTS) methods.

Subsequently, we conduct a \textit{fidelity} assessment to determine if the synthetic evaluation dataset (\texttt{s}) accurately reflects the statistical properties of the real evaluation dataset (\texttt{r}). This assessment employs metrics such as cosine similarity, Euclidean distance, and squared error, which evaluate both the individual feature distributions and the multivariate correlations between features.

Finally, we assess the faithfulness of the synthetic data's utility to the real data's utility using the Wilcoxon signed-rank test \cite{wilcoxon1945individual}. This test operates on paired samples, with two hypotheses: $H_0$, indicating no statistically significant difference between the samples, and $H_1$, indicating a statistically significant difference. The test calculates the absolute differences between paired samples, ranks these differences, and computes a concordance coefficient. The resulting $p$-value is compared to a threshold to determine whether to reject or accept $H_0$. To provide a comprehensive evaluation of classifier parameters, we calculate the average of the $p$-values obtained from the \textit{utility} metrics for these hypothesis tests.

\subsection{The Generative Model}

The core intelligence of MalSynGen lies in its neural network model. Our approach is based on the Conditional Generative Adversarial Networks (cGANs) architecture \cite{mirza2014conditional}, an extension of GANs that incorporates conditional variables to guide both the generation and discrimination of synthetic samples. The model consists of a generator, responsible for synthesizing samples, and a discriminator, tasked with distinguishing between real and generated instances. Both components are trained in an adversarial manner, optimizing their cost functions concurrently to reach an equilibrium where the generator produces samples indistinguishable from real ones.

In the proposed model, the sample labels (benign or malicious) are incorporated into the generator and discriminator through an \textit{embedding} layer. This hidden layer projects that information into a latent space compatible with the model inputs. This mechanism explicitly conditions the generation of samples of the desired class, encouraging the modeling of the underlying probability distribution of each data category.

The generator and discriminator sub-models are composed of sequences of densely connected layers, with variable numbers of neurons, interspersed with dropout layers \cite{creswell2018generative}, used to mitigate \textit{overfitting}. The use of dropout aims to reduce the discriminator's excessive dependence on specific patterns of the training set \cite{Kim2023}, preventing the generator from directly replicating the learned examples instead of capturing the latent distribution of the data. Different values were adopted for the dropout rates of both the generator and the discriminator, depending on the dataset considered.

The nonlinear activation applied to the hidden layers is the Leaky Rectified Linear Unit (LeakyReLU) \cite{maas2013rectifier}. This activation function was chosen for its superior training performance compared to other ReLU functions \cite{radford2015unsupervised}. In the output layers, we use the sigmoid function, ensuring that both the generator outputs and the discriminator predictions remain within the range $[0, 1]$, consistent with the normalization of the input data. We decided not to include normalization layers, a common practice in deep networks, in the final implementation based on empirical experiments. This result is justified by the reduced depth of the architecture and the prior normalization of the input data to the range $[0,1]$, which minimizes the occurrence of problems such as dissipation or explosion of the gradient.

For the training process, the loss function Binary Cross-Entropy was used, which is suitable for problems involving binary classification, such as the problem of distinguishing between real and generated samples. This function is widely used for training adversarial networks.

\begin{figure}[!htp]
    \centering
    \includegraphics[width=0.6\linewidth]{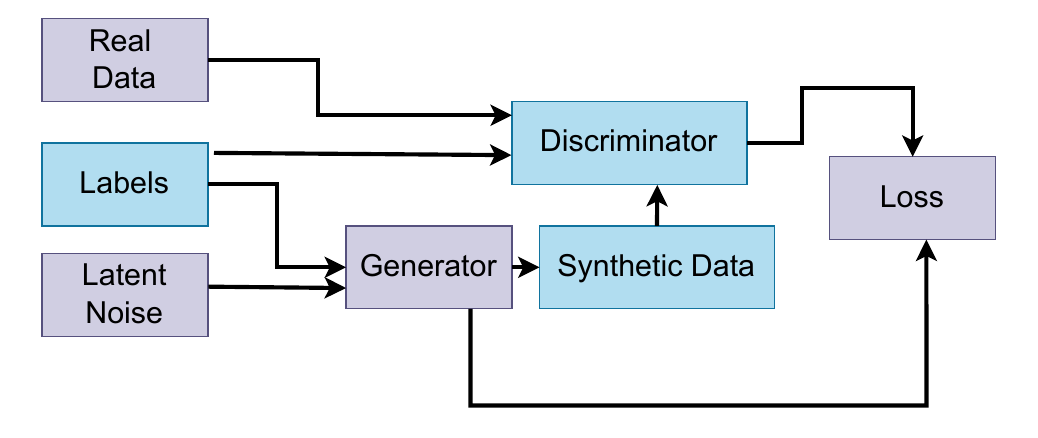}
    \caption{MalSynGen is based on the architecture of Conditional Generative Adversarial Networks (cGANs).}
    \label{fig_gan}
\end{figure}

Figure \ref{fig_gan} illustrates the general architecture of a cGAN, where both the generator and the discriminator are fed labels. The generator uses these labels and latent noise to generate synthetic data, while the discriminator is trained with real data. The discriminator then tries to distinguish between real and synthetic data. Based on the results of the data classification, the loss value of the network is assigned, which is then used to adjust the weights and parameters of the network.

\begin{figure}[!htp]
    \centering
    \includegraphics[width=0.7\linewidth]{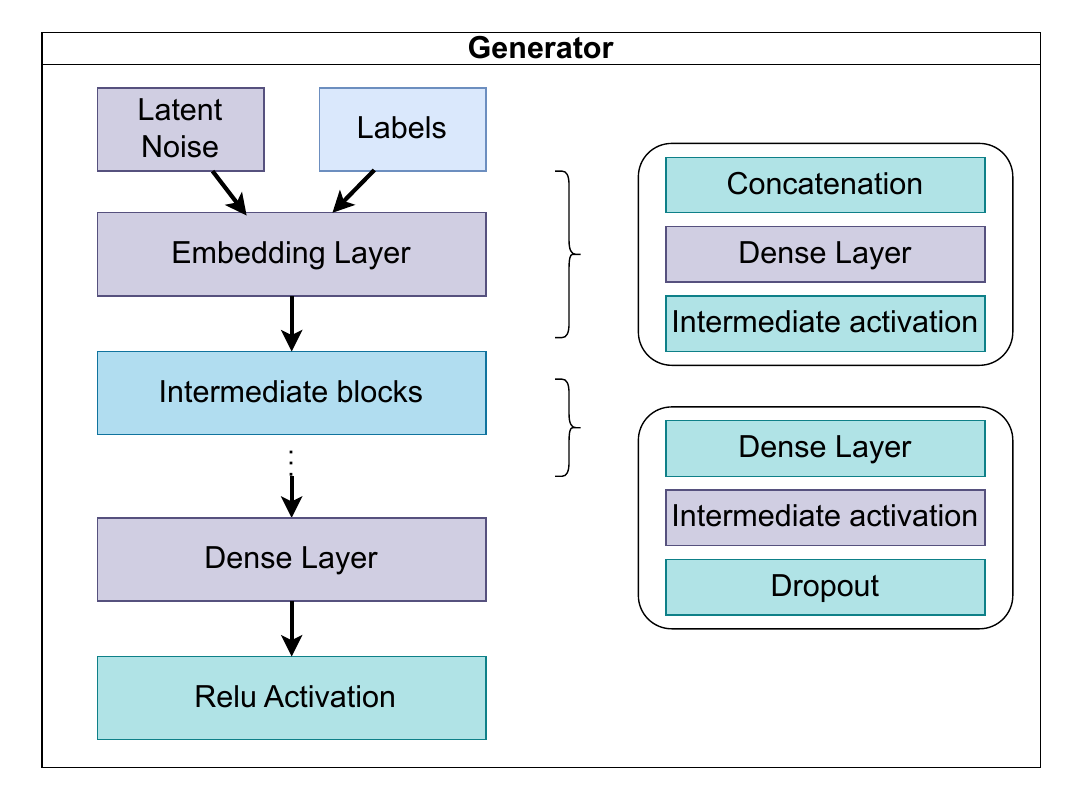}
    \caption{Generator of the cGAN architecture used in MalSynGen.}
    \label{fig_gan2}
\end{figure}

\begin{figure}[!htp]
    \centering
    \includegraphics[width=0.7\linewidth]{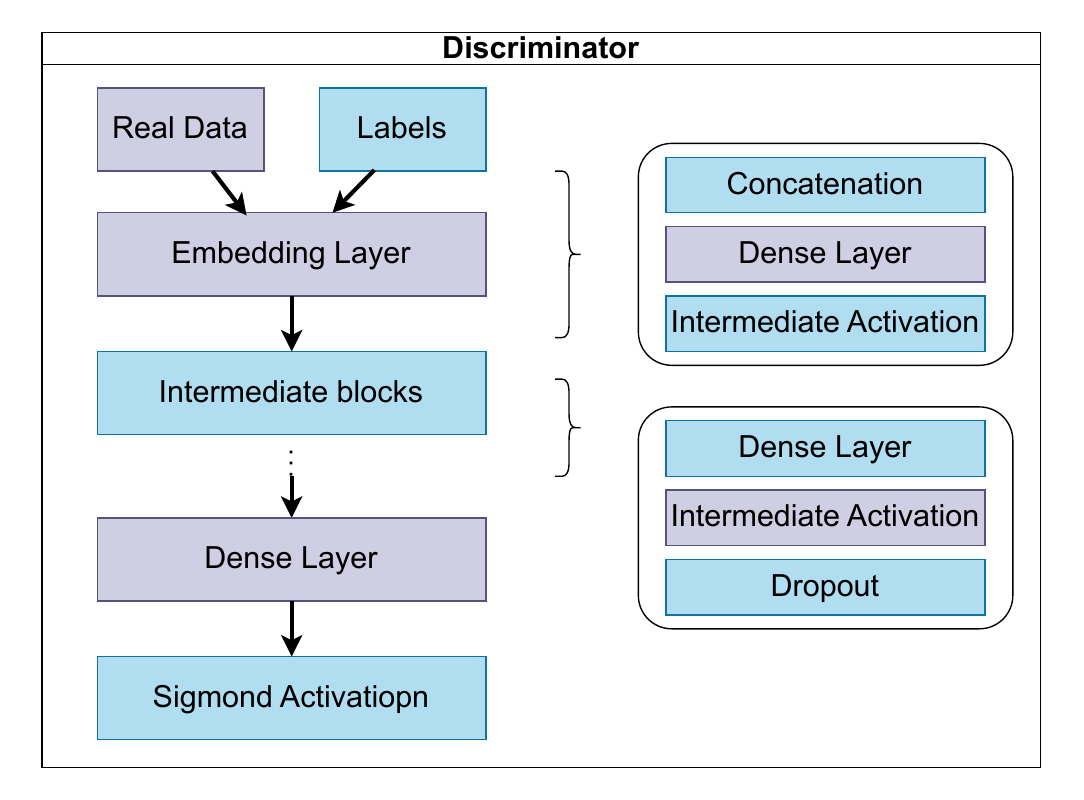}
    \caption{Discriminator of the cGAN architecture used in MalSynGen.}
    \label{fig_gan3}
\end{figure}

Figures \ref{fig_gan2} and \ref{fig_gan3} present the detailed architecture of the cGAN components shown in Figure \ref{fig_gan}: the discriminator and the generator. These components have similar structures, with the main difference being their input and output. Initially, an embedding layer is used to concatenate the latent noise and labels, transforming them into values in the latent space. These values are then processed through intermediate blocks, which consist of dense layers, activation functions, and dropout layers, all aimed at learning the data distribution and performing classification. Finally, dense layers, along with their activation functions, are responsible for normalizing the output and defining the format of the generated data.

\section{Evaluation}
\label{sec:evaluation}

In this section, we present the results of utilizing the MalSynGen framework to generate synthetic datasets. We evaluate two scenarios:
\begin{enumerate}
    \item whether the synthetic dataset preserves the essential characteristics of the original dataset (i.e., fidelity), while effectively extending and diversifying it.
    \item whether the synthetic data can be used with classifiers (i.e., utility).
\end{enumerate}

Table~\ref{tab_hiperparametros} summarizes the hyperparameters used in the instantiation and training of our cGAN model. The evaluation was performed using 10-fold cross-validation ($k$-fold = 10) with a batch size of 256 and the LeakyReLU activation function.

\begin{table*}[!htp]
    \centering 
    \caption{Hyperparameters used in the instantiation and training of our cGAN model.}
    \label{tab_hiperparametros}
    \resizebox{1\textwidth}{!}{%
    \renewcommand{\arraystretch}{1.3} 
        \begin{tabular}{lrrrrrr}
            \hline
            \textbf{Parameter} & \textbf{Androcrawl} & \textbf{Drebin} & \textbf{Adroit} & \textbf{Android P} & \textbf{Kronodroid E} & \textbf{Kronodroid R} \\ \hline
            Epochs & 2,000 & 5,000 & 5,000 & 1,000 & 5,000 & 5,000 \\ \hline
            Neurons per layer (G) & 2,048 & 2,048 & 64 & 1,024 & 512 & 512 \\ \hline
            Neurons per layer (D) & 512 & 1,024 & 32 & 512 & 256 & 256 \\ \hline
            Dropout rate generator & 0.2 & 0.2 & 0.05 & 0.2 & 0.025 & 0.025 \\ \hline
            Dropout rate discriminator & 0.4 & 0.4 & 0.1 & 0.4 & 0.05 & 0.05 \\ \hline
            Initializer deviation & 0.5 & 0.5 & 0.5 & 0.5 & 0.4 & 0.4 \\ \hline
        \end{tabular}%
    }
\end{table*}

The range of epochs was chosen based on the range suggested by \cite{antunes2023hyperparameter}, and the density of neurons per layer was determined based on the ranges explored in \cite{Fristiana2024}, as well as the dropout rates with the supporting middle and upper range from \cite{droputgan2018}, which suggests rates from 0.2 to 0.4 for the dropout rate.

\subsection{Android Malware Datasets}

Table \ref{tab:datasets} presents the specifications of the datasets used in this study, obtained from the Malware-Hunter project public repository\footnote{\url{https://github.com/MalwareDataLab/MalSynGen} on the folder Experiment\_results, accessible in the jbcs tag: git checkout tags/jbcs}. The table provides details on the number of features, as well as the number of malignant and benign samples, including the total number of samples. According to the repository, the datasets include up to 200 features selected using the chi-square method, and each dataset has a balanced total of up to 10,000 samples per class.

\begin{table}[]
\caption{Datasets considered for this study}
\label{tab:datasets}
\centering
\resizebox{0.6\columnwidth}{!}{%
\renewcommand{\arraystretch}{1.3} 
\begin{tabular}{ccccc}
\hline
\textbf{Dataset} & \textbf{Features} & \textbf{Malware} & \textbf{Benign} & \textbf{Total} \\ \hline
\textbf{Adroit} & 118 & 3,418 & 3,418 & 6,836 \\ \hline
\textbf{Androcrawl} & 136 & 10,170 & 10,170 & 20,340 \\ \hline
\textbf{Android Permissions} & 148 & 9,077 & 9,077 & 18,154 \\ \hline
\textbf{Drebin215} & 200 & 5,555 & 5,555 & 11,110 \\ \hline
\textbf{KronoDroid Real Device} & 200 & 10,000 & 10,000 & 20,000 \\ \hline
\textbf{KronoDroid Emulator} & 200 & 10,000 & 10,000 & 20,000 \\ \hline
\end{tabular}%
}
\end{table}

\begin{figure*}[htp]
\centering
\begin{tabular}{cc}
\includegraphics[width=0.5\linewidth]{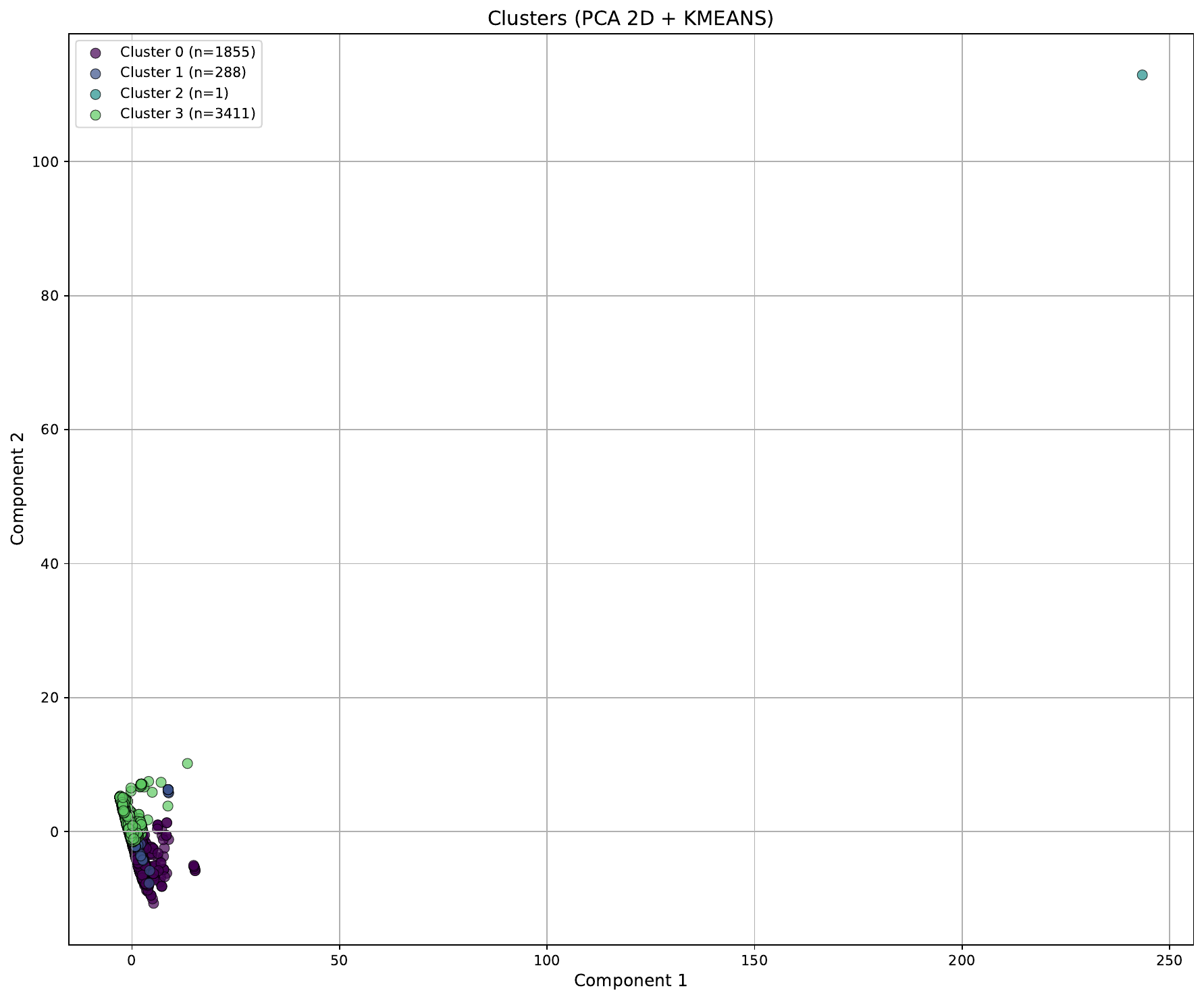} &
\includegraphics[width=0.5\linewidth]{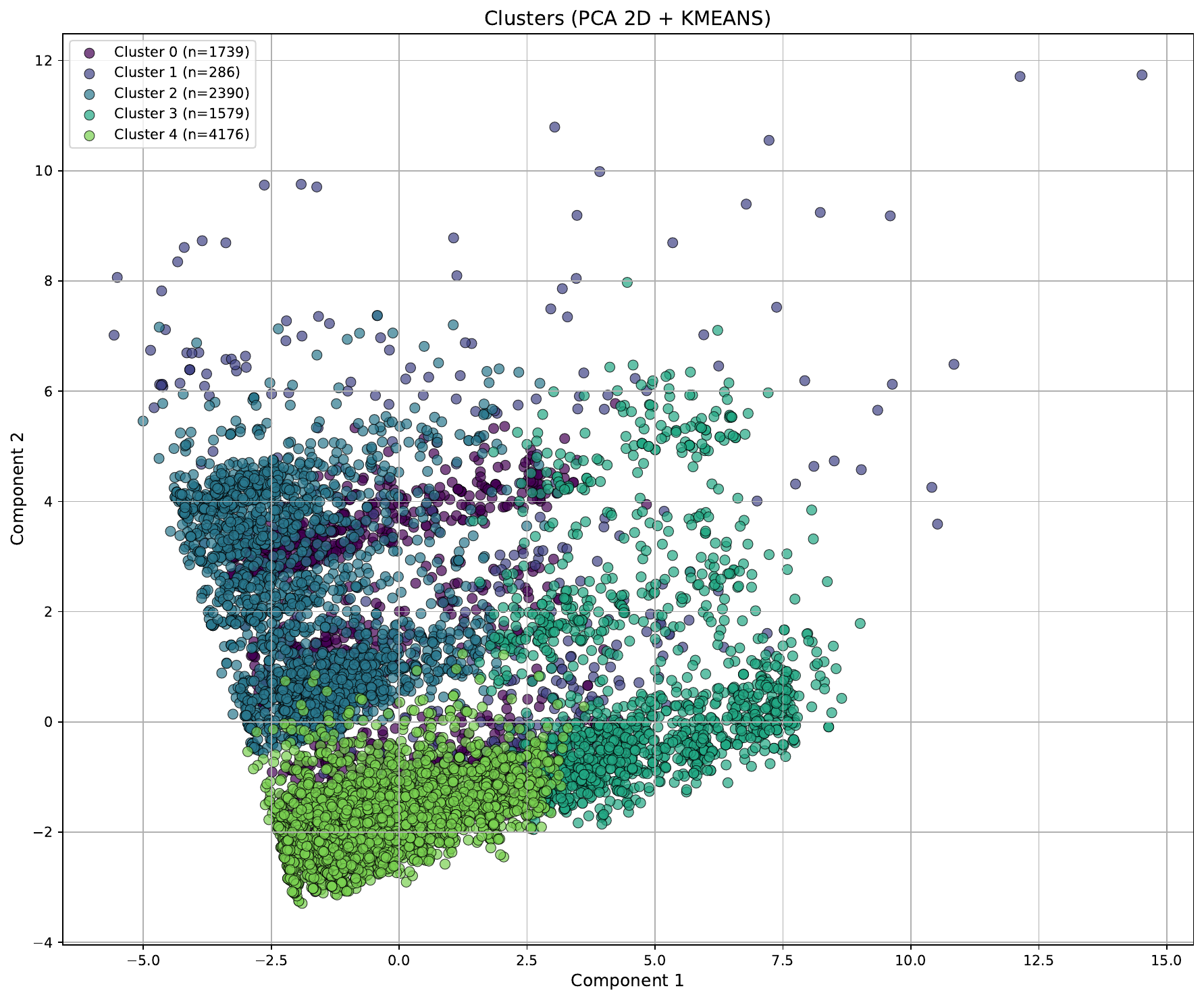} \\
(a) Drebin dataset & (b) Androcrawl dataset \
\end{tabular}
\vspace{+2mm}
\caption{Clustering of malware samples within datasets.}
\label{fig:clustering}
\end{figure*}

In order to further analyze the datasets used, we present in Figure~\ref{fig:clustering} a comparative clustering analysis of malware samples, visualized using PCA-reduced 2D projections with K-means clustering. While clusters are distinguished by color within each plot for visual identification, it is important to note that identical colors across datasets do not indicate cluster correspondence.

The datasets exhibit distinct clustering patterns that indicate the potential existence of malware families, defined as groups of samples that share similar behavioral or structural characteristics. Higher-resolution versions of the clustering visualizations for each dataset are available in the GitHub repository\footnote{\url{https://github.com/MalwareDataLab/MalSynGen} on the folder Datasets,accessible in the jbcs tag: git checkout tags/jbcs}
. Each dataset presents unique clustering dynamics, highlighting variations in malware distribution and feature representation across different data sources.

The Drebin dataset (Figure~\ref{fig:clustering}a) exhibits four clusters. Three of these clusters are tightly grouped in the leftmost corner of the figure. The third cluster (light green) represents the majority of samples (3411), while cluster 2 contains a single outlier sample.

In contrast, the Androcrawl dataset (Figure~\ref{fig:clustering}b) shows more dispersed clustering with five distinct groups. Clusters 0, 2, and 3 maintain somewhat comparable sizes (ranging from 1,500 to 2,300 samples). Cluster 1 (blue) appears to be composed of scattered samples (286), and cluster 4 (light green) emerges as the predominant group (4,176).

Additional plots of malware sample clusters for all other datasets can be found in Appendix~\ref{app:clustering}.
 
\color{black}

\subsection{Fidelity Metrics}
\label{sec_metricas}

In this study, \textit{fidelity} metrics are used to measure the similarity between real and synthetic data. We consider the following \textit{fidelity} metrics: mean squared error, cosine similarity, and squared Euclidean distance, as shown in Table \ref{tab:metrica_fidelidade}, where $n$ represents the total number of samples; $x$ and $y$ are, respectively, the synthetic and real values of each sample.

\begin{table}[h]
    \centering
    \small
    \caption{Fidelity metric formulas used to evaluate the similarity between real data ($x$) and synthetic data ($y$); $n$ represents the total number of samples.}
    \label{tab:metrica_fidelidade}
  \resizebox{0.6\columnwidth}{!}{%
  \renewcommand{\arraystretch}{1.3} 
    \begin{tabular}{@{}lc@{}}
        \toprule
        \textbf{Metric} & \textbf{Formula} \\ \midrule
    
        Cosine similarity & $\displaystyle \frac{x \cdot y}{|x||y|}$ \\ \midrule
        Squared Euclidean distance & $\delta \cdot \delta^T$, where \\
        & $\delta = \mu_{\text{real}} - \mu_{\text{synth}}$, \\
        & $\mu_{\text{real}} = \frac{1}{n} \sum_{i=1}^{n} x_i$, and \\
        & $\mu_{\text{synth}} = \frac{1}{m} \sum_{j=1}^{m} y_j$ \\\midrule
            Mean squared error & $\displaystyle \frac{1}{n}\sum_{i=1}^{n}(x_i - y_i)^2$ \\ 
            \bottomrule
    \end{tabular}
    }
\end{table}

These metrics enable us to verify whether the synthetic dataset reproduces the statistical characteristics of the original dataset. This includes not only the distribution of individual features but also the multivariate correlations between them, ensuring that the synthetic set mirrors the same population as the original. Additionally, we evaluate positive and negative samples separately to ensure that each class closely resembles the original.

Specifically, the Mean Squared Error is a metric used to assess the quality of a regression model. It is measured by calculating the mean of the squared differences between the predicted values (\(x_i\)) and the real values (\(y_i\)). A lower value indicates that the predicted values are closer to the real values. However, a value of zero is not ideal in our case, as it indicates that the model generated identical values to the real values.

Cosine similarity is a metric used to measure the similarity between two vectors. In our case, it is used to compare the synthetic and real values (\(x\) and \(y\), respectively). The similarity value ranges between 0 and 1, with values closer to 1 indicating higher similarity between the vectors. However, a value of 1 is not desirable, as it indicates that both sets are identical.

The squared Euclidean distance, similarly to the mean squared error, is a metric used to measure the difference between two sets. It returns the squared difference between the means of the real and synthetic samples. A smaller value indicates that the predicted values are closer to the real values. However, a value of zero indicates that both sets are identical; therefore, an ideal value approaches zero but is not equal to zero.

\subsection{Utility Metrics}

We also consider \textit{utility} metrics to assess the performance of machine learning classifiers trained with synthetic data, considering both TRTS (train on real, test on synthetic) and TSTR (train on synthetic, test on real) evaluation methods. To evaluate the utility of the generated data, we considered four classifiers: \textit{SupportVectorMachine} (SVM), \textit{Decision Tree} (DT), \textit{XGBoost}, and \textit{Stochastic Gradient Descent Regressor} (SGDR).

In each fold, we applied the classifiers to both the real and synthetic datasets generated by each cGAN configuration. Based on the results, we generated confusion matrices and ROC curves, and extracted the previously defined \textit{utility} metrics: accuracy, precision, recall, F1-score, and AUC (Area Under the ROC Curve).

While accuracy, precision, recall, and F1-score are commonly used metrics, we also focus on the AUC, a statistical metric used to evaluate the performance of a machine learning model, particularly in binary classification tasks. The AUC graphically represents the relationship between the true positive rate (TPR) and the false positive rate (FPR) across different decision thresholds. A value of 1 indicates a model with perfect predictions, 0.5 indicates a model that makes random decisions, and a value of 0 suggests the model makes all predictions incorrectly. The formula for calculating the AUC is as follows:

\begin{equation} 
    AUC = TPR - (1 - \frac{TN}{FP + TN}) \cdot FPR
    \label{eq:auc}
\end{equation}

To verify the statistical significance of the metrics, we utilize the Mann-Whitney U test. The Mann-Whitney U test operates on two independent samples with two hypotheses: $H_0$, which states that there is no statistically significant difference between the distributions of the samples, and $H_1$, which states that there is a statistically significant difference. This is determined by calculating the $U$ statistic based on the sum of the ranks of ordered observations from the same sample. This calculation results in a $p$-value, which is compared with a threshold to determine whether to reject or accept $H_0$.

To complement and validate the utility metrics, we calculate the average of the $p$-values obtained from the \textit{utility} metrics for our null hypothesis ($H_0$), which posits that there is no statistically significant difference between the performance of the two classifier sets, thereby demonstrating comparable utility. This averaging provides a comprehensive assessment of the classifier parameters. Furthermore, we adopt a threshold of 0.05 for the $p$-value, a standard value widely accepted in scientific literature, to conduct our experiments.

In addition to the utility metrics, we also collect data on the consumption of computational resources at the end of the evaluation. This includes the percentage of CPU and memory usage during each iteration of the training stage, providing insights into the computational demands of our approach.

\subsection{Environment and Technologies}
\label{sub:enviroment}

Four machines with Debian GNU/Linux 12 distributions (kernel 6.1.0-27-amd64), x86-64 architectures, Intel(R) Core(TM) i7-9700 CPUs, and 16 GB of RAM were used to perform the experiments in this study.

To implement and execute the core architecture of our cGAN, we used Python (version 3.8) and the following main libraries:

\begin{itemize}
    \item \textit{NumPy} 1.21.5: Used for various vector and matrix operations involving the labels and samples generated by the network.
    \item \textit{Keras} 2.9.0: Used to define the conditional models of the generator and discriminator.
    \item \textit{TensorFlow} 2.9.1: Applied in conjunction with Keras to define the network's loss functions.
    \item \textit{Pandas} 1.4.4: Responsible for processing and handling input and output data in CSV format.
    \item \textit{scikit-learn} 1.1.1: Used to stratify folds during executions and to calculate evaluation metrics.
    \item \textit{MLflow}\footnote{\url{https://mlflow.org/}}: To track computational resource consumption metrics.
\end{itemize}

Furthermore, more details on the cGAN implementation, resources used, and results of all executions can be found in the GitHub repository\footnote{\url{https://github.com/MalwareDataLab/
MalSynGen}}.

\section{Results and discussion}\label{sec:results}


\subsection{Utility metrics}

\begin{figure*}[htp]
    \centering
 
    \includegraphics[width=1\linewidth]{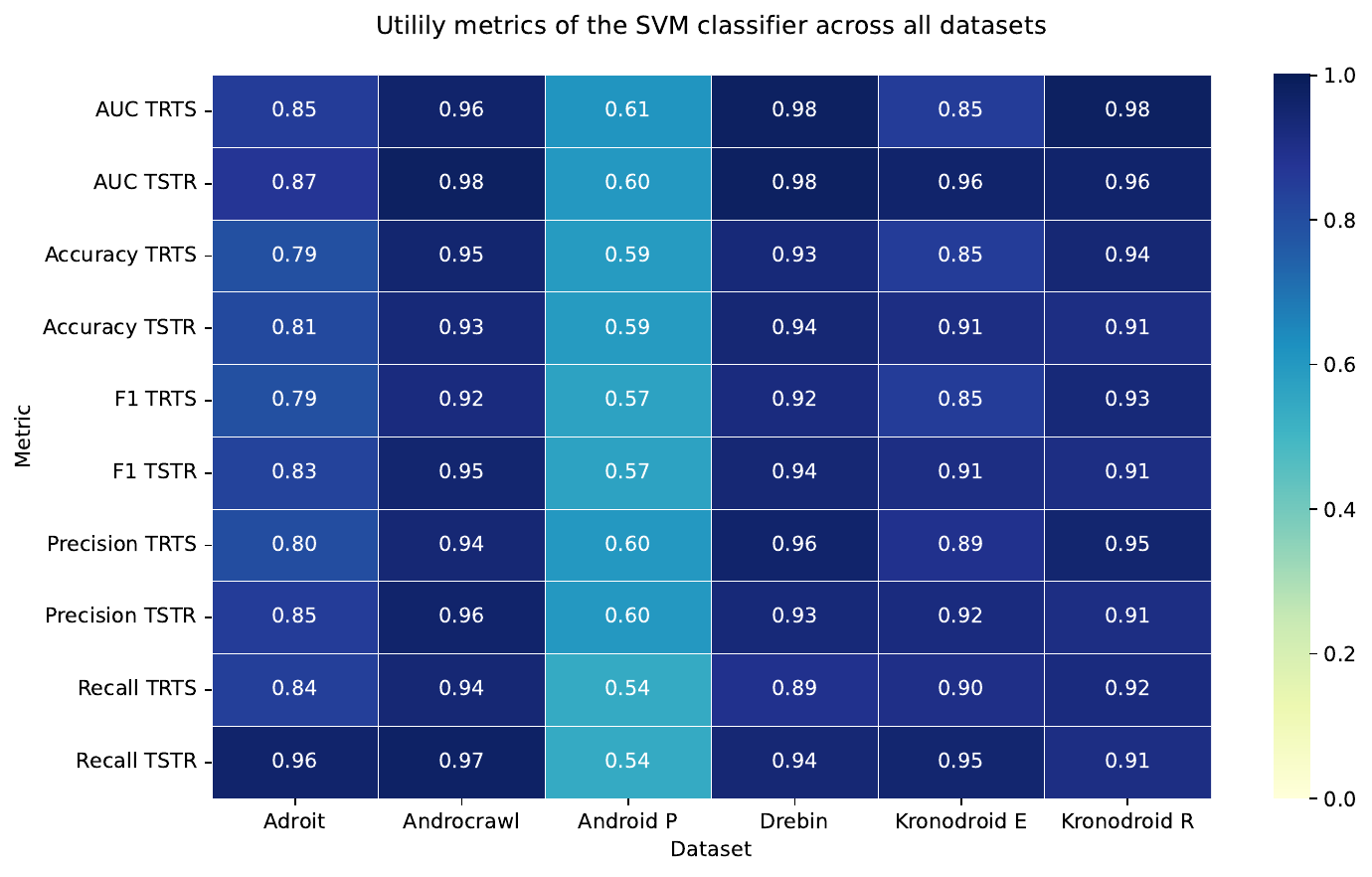} 
    \vspace{+2mm}
    \caption{Performance heatmap of SVM classifier utility metrics across all datasets for TSTR and TRTS scenarios.}
    \label{fig:utilidade_all_datasets}
\end{figure*}

Figure~\ref{fig:utilidade_all_datasets} presents the average 10-fold cross-validation results for the SVM classifier across all six datasets, visualized through a performance heatmap. In this visualization, darker shading indicates better performance (values closer to 1.0), while lighter shading signifies poorer performance (values closer to 0.0). For each of the six datasets, the heatmap organizes results by five key evaluation metrics (accuracy, precision, F1-score, AUC, and recall), with each metric showing paired rows for both TSTR (Train on Synthetic, Test on Real) and TRTS (Train on Real, Test on Synthetic) scenarios.

With the sole exception of the Android Permission dataset, the remaining datasets consistently demonstrated high utility metrics, with values ranging from 0.79 to 0.98 across all evaluated metrics in both TSTR and TRTS scenarios. Particularly noteworthy are the recall values, which ranged from 0.84 to 0.97. This is a crucial metric for malware classification, as false negatives carry significant security implications. These robust results strongly suggest that the generated synthetic data is highly useful for malware classification across multiple diverse datasets. Additionally, the diminished performance observed for the Android Permission dataset (with metrics ranging between 0.55 and 0.61) likely reflects inherent constraints in its source data. This is further supported by prior experiments using real-world data for both evaluation and training, which yielded a similarly low performance for this specific dataset.

A complete list of all classifiers used in this study, along with their corresponding utility metrics, can be found in Appendix~\ref{app:utility_metrics}.
\color{black}

In Table \ref{tab_valores_de_p}, we present the $p$-values obtained from the Wilcoxon test for accuracy, precision, F1-score, recall, and AUC for each classifier. Each $p$-value indicates the statistical significance of the differences between the classifier sets.

\begin{table}[h!]
    \centering
    \small
    \caption{$p$-values of the classifier metrics}
    \label{tab_valores_de_p}
    \resizebox{0.7\columnwidth}{!}{%
    \renewcommand{\arraystretch}{1.3} 
        \begin{tabular}{ccccc}
            \hline
            \textbf{Dataset} & \textbf{SVM} & \textbf{Decision Tree} & \textbf{SGD Regressor} & \textbf{XGBoost} \\ \hline
            Adroit & 0.0668 & 0.3906 & 0.2254 & 0.1563 \\ \hline
            Androcrawl & 0.0570 & 0.1566 & 0.4945 & 0.1281 \\ \hline
            Android Permissions & 0.4287 & 0.3020 & 0.5871 & 0.7745 \\ \hline
            Drebin & 0.1801 & 0.3223 & 0.2164 & 0.3219 \\ \hline
            Kronodroid Emulator & 0.1134 & 0.0684 & 0.1062 & 0.0348 \\ \hline
            Kronodroid Real & 0.1109 & 0.3641 & 0.0602 & 0.1867 \\ \hline
        \end{tabular}%
    }
\end{table}







The results indicate that in most cases, there is no statistically significant difference between the classifier sets, regardless of the classifier used. The only exception was observed with the XGBoost classifier for the Kronodroid Emulator dataset, which had a $p$-value of 0.0348. Although this value is below the defined threshold, it is close to the limit. Even so, the model's performance metrics remain high, with accuracy ranging from 0.86 to 0.88, recall between 0.85 and 0.88, and a high AUC, ranging from 0.93 to 0.95. Furthermore, an anomalous behavior is observed with the Android Permission dataset, where the $p$-values are high, between 0.3020 and 0.7745, indicating high similarity between the performance of the two classifier sets, despite the low utility metrics.

\begin{figure*}[htp]
    \centering
        \includegraphics[width=0.85\linewidth]{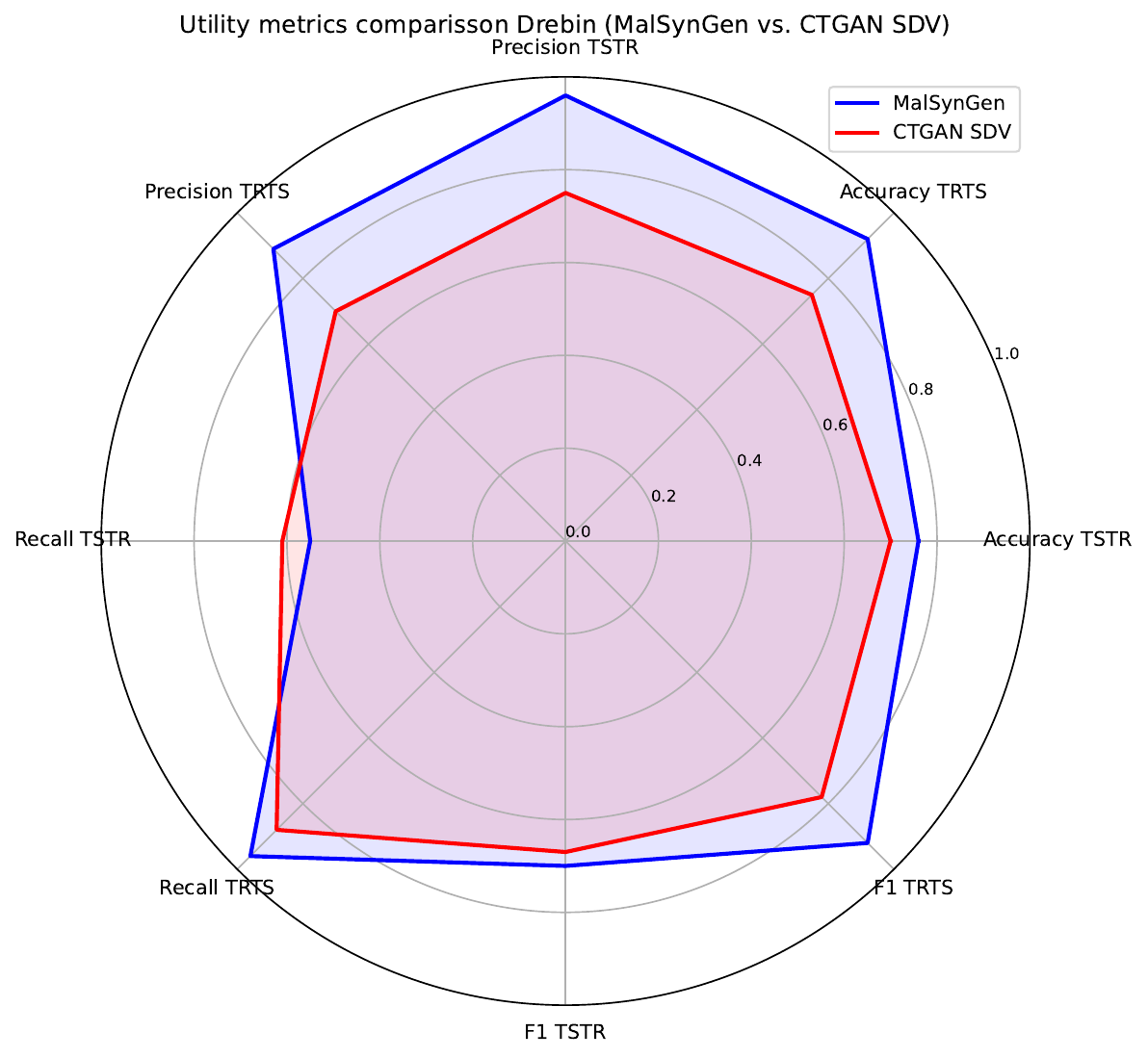} 
\caption{Comparative utility metrics for MalSynGen and CTGAN  SVM performance on Drebin dataset}
    \label{fig:utilidade_drebin}
\end{figure*}

\begin{figure*}[htp]
    \centering
        \includegraphics[width=0.85\linewidth]{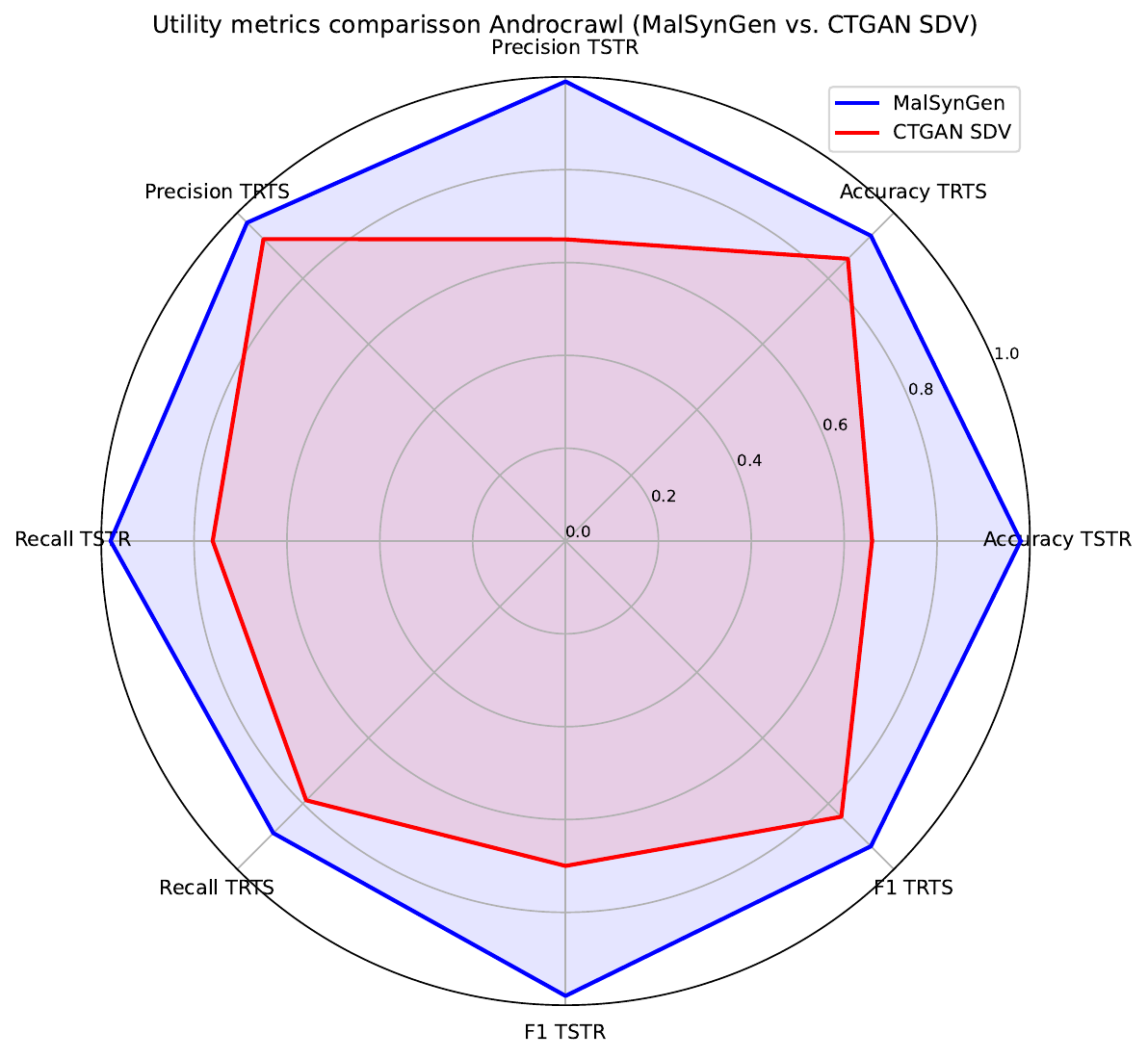} 
\caption{Comparative utility metrics for MalSynGen and CTGAN  SVM performance on Androcrawl dataset}
 
    \label{fig:utilidade_androcrawl}
\end{figure*}

In Figures~\ref{fig:utilidade_drebin} and~\ref{fig:utilidade_androcrawl} we present a comparative analysis of MalSynGen against SDV's CTGAN \cite{xu2019modeling}. This comparison is conducted through 10-fold cross-validation using SVM classifiers on the Drebin and Androcrawl datasets.

For both MalSynGen and CTGAN, identical hyperparameters are maintained for the AndroCrawl and Drebin datasets, as specified in Table~\ref{tab_hiperparametros}. The only exceptions are a reduced layer configuration (256G/64D layers) and the application of 100 training epochs across both models. The evaluation comprehensively implements both TSTR (Train on Synthetic, Test on Real) and TRTS (Train on Real, Test on Synthetic) methodologies for each model.
 

The results indicate that MalSynGen generates synthetic data with superior utility metrics across both evaluation paradigms (TSTR and TRTS), demonstrating an average improvement of 0.18375 on the Androcrawl dataset and 0.1025 on the Drebin dataset, compared to CTGAN. Although a slight exception is observed in the TSTR recall value for the Drebin dataset (0.55 for MalSynGen versus 0.61 for CTGAN), the overall findings suggest that MalSynGen consistently outperforms CTGAN in the context of Android malware classification for this specific scenario.

\color{black}
\subsection{Fidelity metrics}

In Table~\ref{tab_fidelidade_positivo} and Table~\ref{tab_fidelidade_negativo}, we present the results obtained for the \textit{fidelity} metrics (cosine similarity,  squared Euclidean distance, and mean squared error) for the datasets considered in this study.

\begin{table}[h]
    \centering
    \small
    \caption{Fidelity metric values for positive samples (malware).}
    \label{tab_fidelidade_positivo}
     \resizebox{0.6\columnwidth}{!}{%
     \renewcommand{\arraystretch}{1.3} 
    \begin{tabular}{@{}cccc@{}}
        \toprule
        \multicolumn{4}{c}{\textbf{Positive Samples}} \\ \midrule
        \textbf{Dataset} & \textbf{Cosine} & \textbf{Euclidean Distance} & \textbf{Squared Error} \\ \midrule
        Adroit & 0.697 & 0.097 & 0.052 \\ \hline
        Androcrawl & 0.602 & 0.151 & 0.116 \\ \hline
        Android P & 0.321 & 0.017 & 0.035 \\ \hline
        Drebin & 0.370 & 0.398 & 0.138 \\ \hline
        Kronodroid E & 0.632 & 0.352 & 0.141 \\ \hline
        Kronodroid R & 0.626 & 0.277 & 0.167 \\ \hline
    \end{tabular}
    }
\end{table}

\begin{table}[h]
    \centering
    \small
    \caption{Fidelity metric values for negative samples (benign).}
    \label{tab_fidelidade_negativo}
     \resizebox{0.6\columnwidth}{!}{%
     \renewcommand{\arraystretch}{1.3} 
    \begin{tabular}{cccc}
        \hline
        \multicolumn{4}{c}{\textbf{Negative Samples}} \\ \hline
        \textbf{Dataset} & \textbf{Cosine} & \textbf{Euclidean Distance} & \textbf{Squared Error} \\ \hline
        Adroit & 0.589 & 0.127 & 0.045 \\ \hline
        Androcrawl & 0.523 & 0.050 & 0.088 \\ \hline
        Android P & 0.359 & 0.016 & 0.037 \\ \hline
        Drebin & 0.540 & 0.262 & 0.169 \\ \hline
        Kronodroid E & 0.676 & 0.110 & 0.098 \\ \hline
        Kronodroid R & 0.641 & 0.102 & 0.104 \\ \hline
    \end{tabular}
    }
\end{table}

The cosine similarity results indicate that the synthetic data is similar, but not identical, to the original data, as the values approach 1, except for the Android P and Drebin datasets. However, as previously emphasized, the metrics obtained from classifiers trained with Android P are inherently low, leading to the synthetic data exhibiting similar behavior. Regarding Drebin, despite having a cosine similarity of 0.370 for positive samples, which is low compared to other results, its other metrics and the value for negative samples indicate that the data still capture the data patterns.

The mean squared error values indicate high similarity between the synthetic and original data for all datasets, as they are close to zero, though not exactly zero. Furthermore, the Euclidean distance values corroborate the squared error results. The only exceptions are the Drebin, Kronodroid E, and Kronodroid R datasets, which show slightly higher values. Nevertheless, these values remain within an acceptable threshold, and the other metrics demonstrate positive results.

\subsection{Computational Resources}

In Table~\ref{tab_consumo_recursos}, we present the results obtained by tracking the consumption of computational resources (CPU and memory), along with the total time used to run all experiments (i.e., 10 runs for TRTS and TSTR each). The experiments were conducted in the environment described in Section \ref{sub:enviroment}.

\begin{table}[h]
    \centering
    \small
    \caption{Average consumption of computational resources.}
    \label{tab_consumo_recursos}
     \resizebox{0.6\columnwidth}{!}{%
     \renewcommand{\arraystretch}{1.3} 
    \begin{tabular}{cccc}
        \hline
        \multirow{2}{*}{\textbf{Dataset}} & \multirow{2}{*}{\textbf{CPU (\%)}} & \multirow{2}{*}{\textbf{Memory (\%)}} & \multirow{2}{*}{\textbf{Execution time}} \\
        & & & \\ \hline
        Adroit & 44.09\% & 15.51\% & 0.731h \\ \hline
        Androcrawl & 75.49\% & 17.18\% & 2.8h \\ \hline
        Android P & 66.04\% & 23.87\% & 2.2h \\ \hline
        Drebin & 78.40\% & 18.10\% & 4.1h \\ \hline
        Kronodroid E & 50.97\% & 18.33\% & 3.4h \\ \hline
        Kronodroid R & 51.97\% & 18.10\% & 3.3h \\ \hline
    \end{tabular}
    }
\end{table}

The results suggest that configurations with higher layer density (Androcrawl and Drebin) exhibit the highest CPU usage, while those with a greater number of epochs (Adroit, Kronodroid E, and Kronodroid R) show the highest memory consumption and execution time. This indicates a correlation between hyperparameters and computational resource consumption. Additionally, the results demonstrate that MalSynGen is CPU-intensive but does not consume excessive memory during execution. The variations in computational resource consumption are primarily influenced by dataset size and hyperparameter values.

\section{Final Remarks}\label{sec:conclusion}

MalSynGen is a publicly available framework designed for training and evaluating cGAN generative networks, specifically for generating synthetic tabular datasets in the context of Android malware detection. It employs a methodology that rigorously assesses both the utility and fidelity of the generated data.

Key findings from experiments conducted on six Android malware datasets demonstrate that the synthetic datasets produced by MalSynGen are:

\begin{itemize}
    \item \textbf{Useful}: Exhibiting high utility metrics, with values ranging from 0.75 to 0.98 in five datasets, and high AUC scores (0.80 to 0.99), confirming consistent and positive performance across various classifiers.
    \item \textbf{Faithful}: Maintaining fidelity to the original data, as evidenced by cosine similarity approaching 1 in four datasets, mean squared error values close to zero, and Euclidean distance values within acceptable thresholds.
\end{itemize}

Analysis of computational resource consumption indicates that MalSynGen is CPU-intensive but does not require excessive memory. Furthermore:

\begin{itemize}
    \item Configurations with higher layer densities increase CPU usage.
    \item Configurations with a higher number of epochs demand more memory and execution time.
\end{itemize}

Future research directions may include:

\begin{itemize}
    \item Expanding the evaluation metrics to include additional measures of utility, fidelity, and computational performance, and exploring new classifiers.
    \item Conducting a comparative performance analysis with  general purpose synthetic data generation tools using identical datasets.
    \item Generalizing the methodology to incorporate other neural network models for synthetic tabular data generation.
    \item Evaluating the framework in the context of diverse cybersecurity threats beyond Android malware.
\end{itemize}

\section*{Declarations}

\subsection*{Contributions}
Conceptualization: All authors contributed to the conceptualization of this research.
Methodology: All authors were involved in the development of the methodology.
Software: Kayua Oleques Paim, Hendrio Bragança, and Angelo Gaspar were responsible for software development.
Validation: All authors participated in the validation process.
Formal Analysis: All authors contributed to the formal analysis of the results.
Investigation: All authors were involved in the investigation.
Resources: Diego Kreutz and Rodrigo Mansilha provided resources for this research.
Data Curation: All authors were involved in data curation.
Writing - Original Draft Preparation: All authors contributed to the original draft preparation.
Writing - Review and Editing: All authors reviewed and edited the manuscript.
Supervision: Diego Kreutz and Rodrigo Mansilha provided supervision for this research.
Project Administration: Diego Kreutz and Rodrigo Mansilha administered the project.
Funding Acquisition: Diego Kreutz and Rodrigo Mansilha secured funding for this research.

\subsection*{Interests}
The authors declare that they have no competing interests.

\subsection*{Acknowledgements}
The authors extend their gratitude to Anna Luiza Gomes da Silva and Lucas Ferreira Areias de Oliveira for their invaluable contributions to hyperparameter configuration and resource collection. The authors also express their sincere appreciation to the anonymous SBSeg 2024 reviewers for their insightful suggestions and feedback.

\subsection*{Funding}
This research was partially supported by the National Education and Research Network (RNP) through the ``Programa Hackers do Bem'' initiative and the GT Malware DataLab. It also received support from the Coordination for the Improvement of Higher Education Personnel (CAPES), Brazil, Financing Code 001, and from the Research Support Foundation of the State of Rio Grande do Sul (FAPERGS) under grant agreements 2/2551-0000841-0, 24/2551-0001368-7, and 24/2551-0000726-1.

\subsection*{Materials}
Code: \url{https://github.com/MalwareDataLab/MalSynGen}\\
Datasets: The datasets can be found in the \texttt{datasets} directory located inside the GitHub repository.

\bibliographystyle{ieeetr}
\bibliography{refs}

\newpage

\begin{appendices}

\section{Classifier Utility Metrics Across All Datasets}
\label{app:utility_metrics}

In this appendix, we present the set of utility metrics for all classifiers across each evaluated dataset. We detail the mean values of the utility metrics (AUC, precision, recall, accuracy, and F1-score) obtained for MalSynGen through a 10-fold evaluation process, employing both the TSTR (Train on Synthetic, Test on Real) and TRTS (Train on Real, Test on Synthetic) methods.

\begin{figure}[!htbp]
    \centering
    \includegraphics[width=\linewidth]{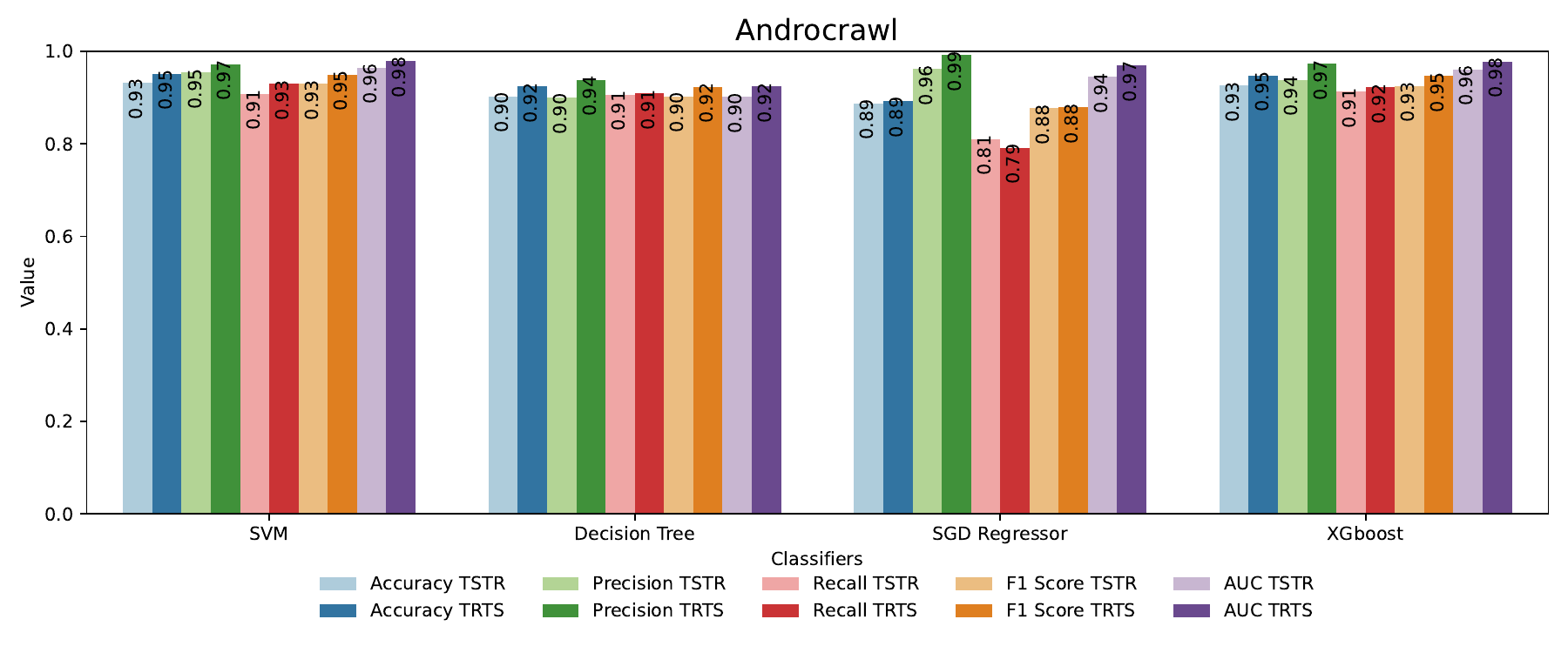}
    \caption{Utility metrics AndroCrawl dataset}
    \label{fig:utilidade_androcrawl_bar}
\end{figure}

\begin{figure}[!h]
    \centering
    \includegraphics[width=\linewidth]{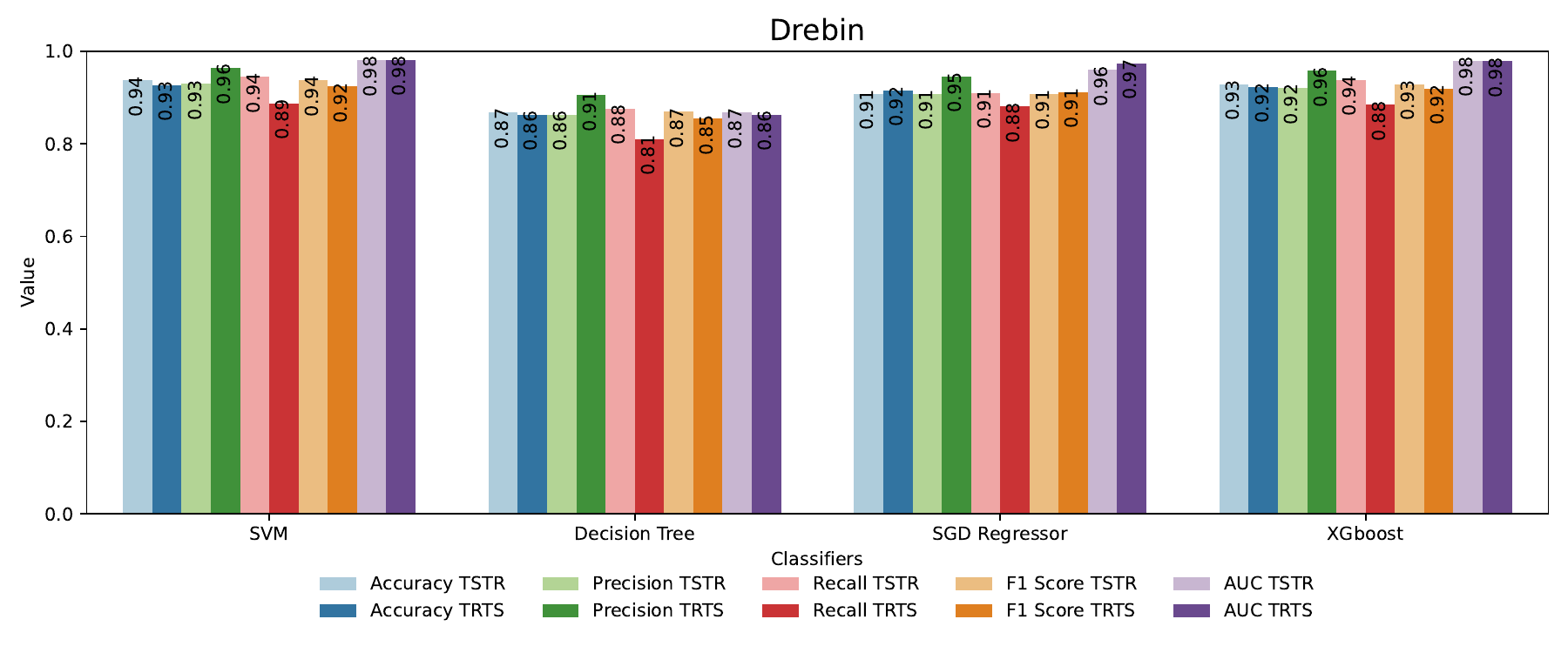}
    \caption{Utility metrics Drebin dataset}
    \label{fig:utilidade_drebin_bar}
\end{figure}

\begin{figure}[!htbp]
    \centering
    \includegraphics[width=\linewidth]{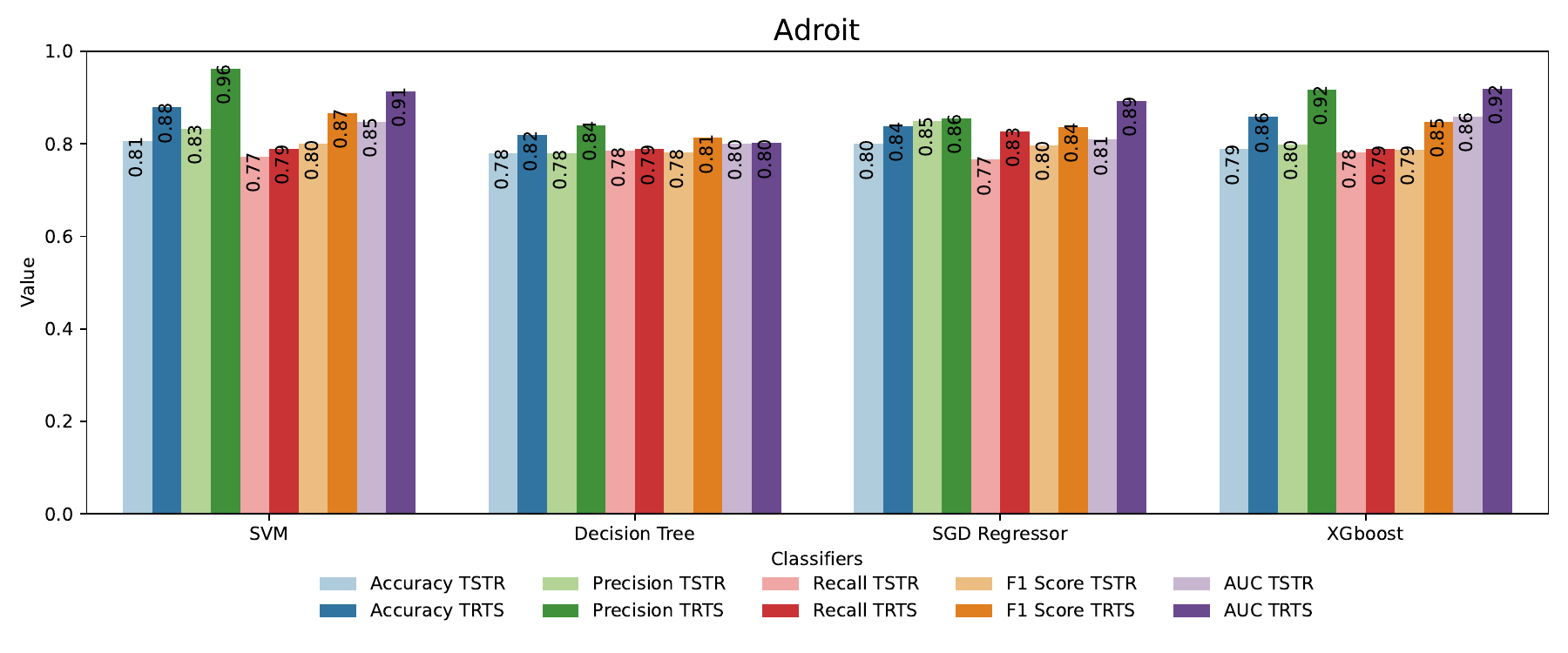}
    \caption{Utility metrics Adroit dataset}
    \label{fig:utilidade_androit_bar}
\end{figure}

\begin{figure}[!htbp]
    \centering
    \includegraphics[width=\linewidth]{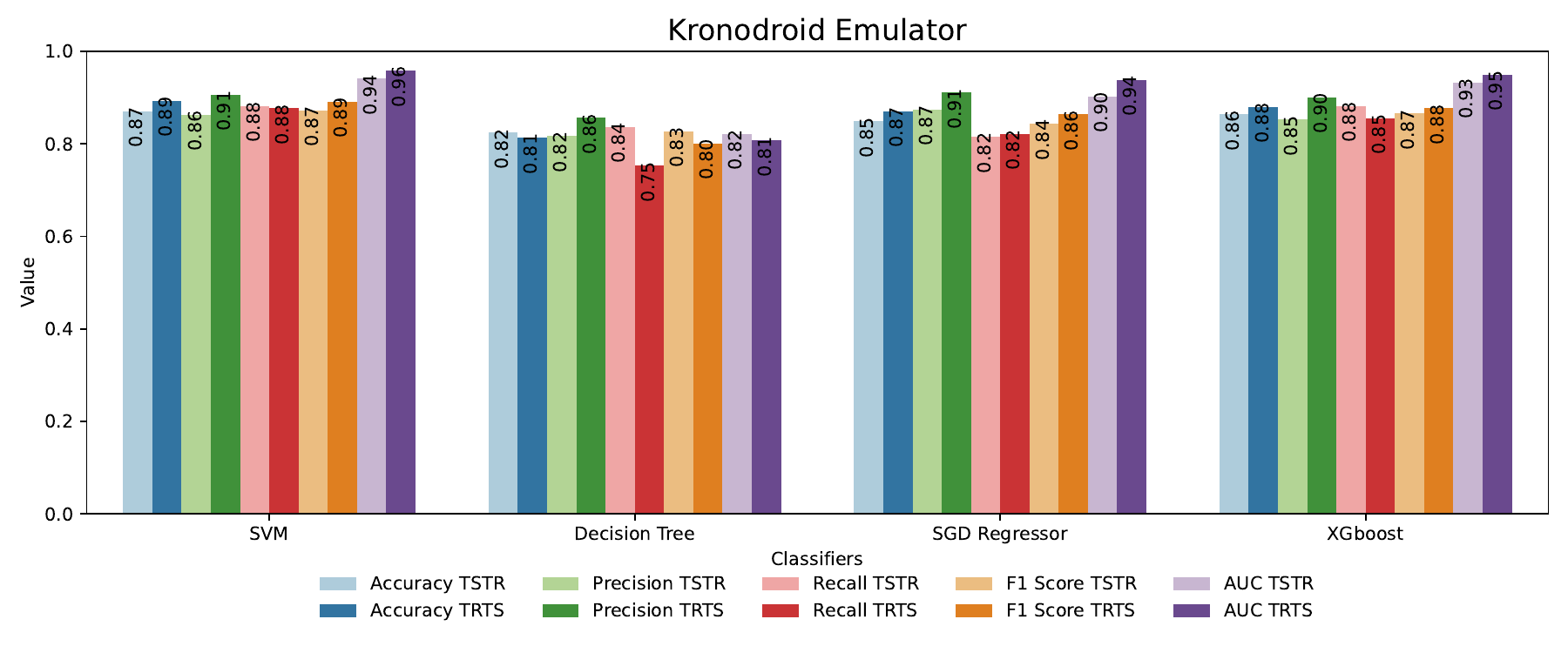}
    \caption{Utility metrics Kronodroid E dataset}
    \label{fig:utilidade_kronodroid_emu_bar}
\end{figure}

\begin{figure}[!htbp]
    \centering
    \includegraphics[width=\linewidth]{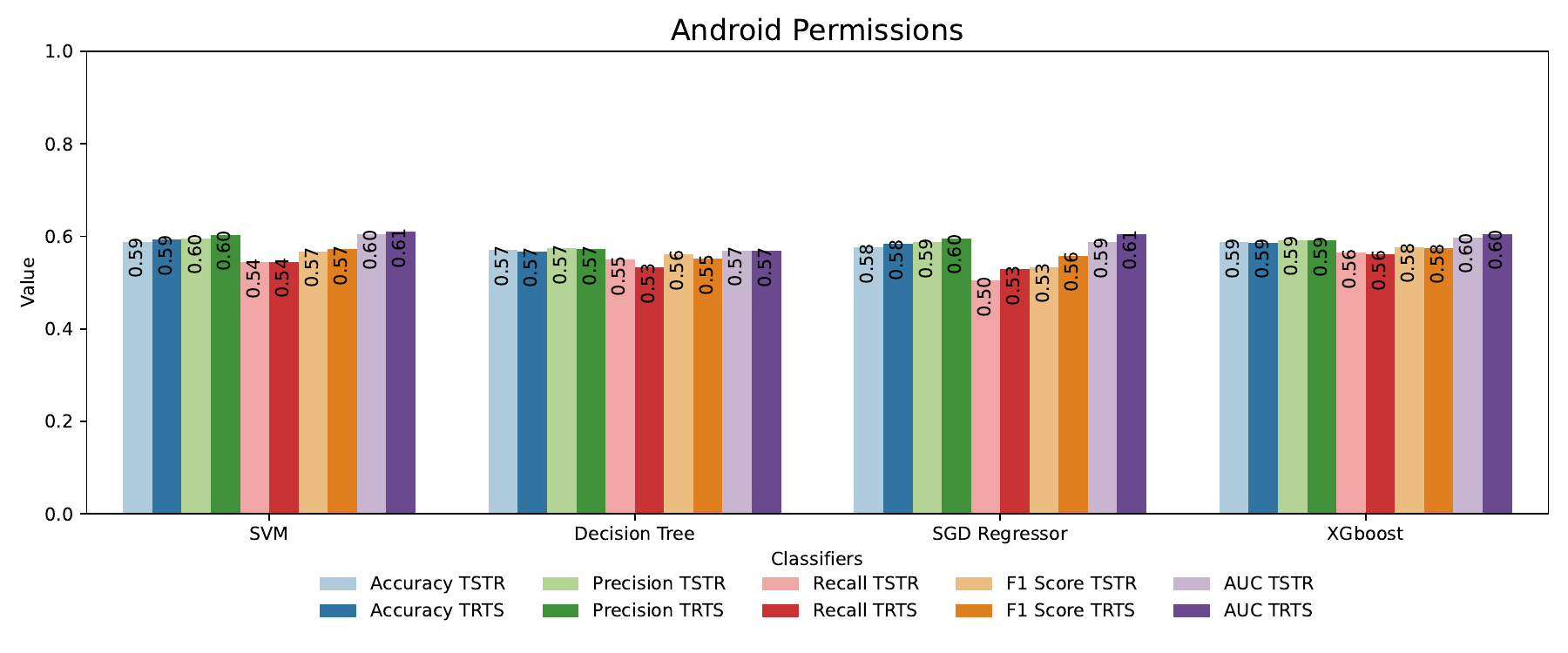}
    \caption{Utility metrics Android P dataset}
    \label{fig:utilidade_android_permission_bar}
\end{figure}

\begin{figure}[!htbp]
    \centering
    \includegraphics[width=\linewidth]{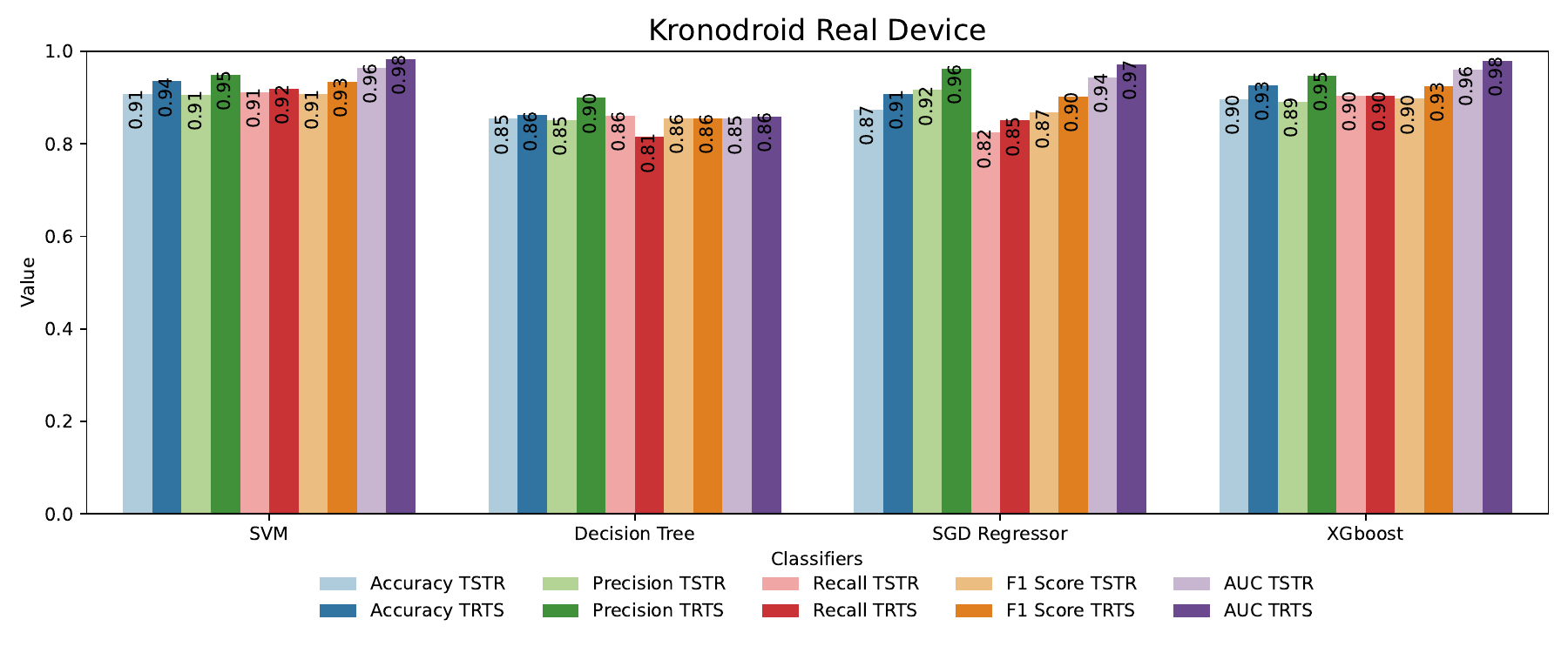}
    \caption{Utility metrics Kronodroid R dataset}
    \label{fig:utilidade_kronodroid_real_bar}
\end{figure}

The detailed results are visualized in:
\begin{itemize}
    \item Figure~\ref{fig:utilidade_androcrawl_bar} (Androcrawl) \footnote{Higher resolution of the images can be found at \url{https://github.com/MalwareDataLab/Datasets-JBCS/tree/b53e8a184003d775e02c2dc3e9b953ffcb8d6a9c/Figures}}

    \item Figure~\ref{fig:utilidade_drebin_bar} (Drebin)
    \item Figure~\ref{fig:utilidade_androit_bar} (Adroit)
    \item Figure~\ref{fig:utilidade_kronodroid_emu_bar} (Kronodroid Emulator)
    \item Figure~\ref{fig:utilidade_android_permission_bar} (Android Permission)
    \item Figure~\ref{fig:utilidade_kronodroid_real_bar} (Kronodroid Real Device)
\end{itemize}
For each classifier, these figures display five pairs of bars, with each pair representing one of the five metrics (accuracy, precision, F1-score, AUC, and recall). Within each pair, one bar illustrates the value for the TSTR scenario, and the other for the TRTS scenario.

Notably, all classifiers consistently exhibited high utility (0.75--0.9), particularly in AUC, precision, and F1-score, across most datasets. As previously discussed, the Android Permissions dataset represented an exception. These consistent results further confirm the practical utility of our synthetic data for malware classification. We also observed a stable performance hierarchy among the classifiers: Support Vector Machines (SVM) consistently delivered the highest utility, followed by XGboost, SGD Regressor, and Decision Tree models.

\section{Malware Sample Clustering Across All Datasets}
\label{app:clustering}

In this appendix, we present the clustering results for the other datasets considered in this study, visualized using PCA-reduced 2D projections with K-means clustering. While we distinguish clusters by color within each plot for visual identification, we note that identical colors across datasets do not indicate cluster correspondence.

\begin{figure}[!htbp]

        \centering
        \includegraphics[width=\linewidth]{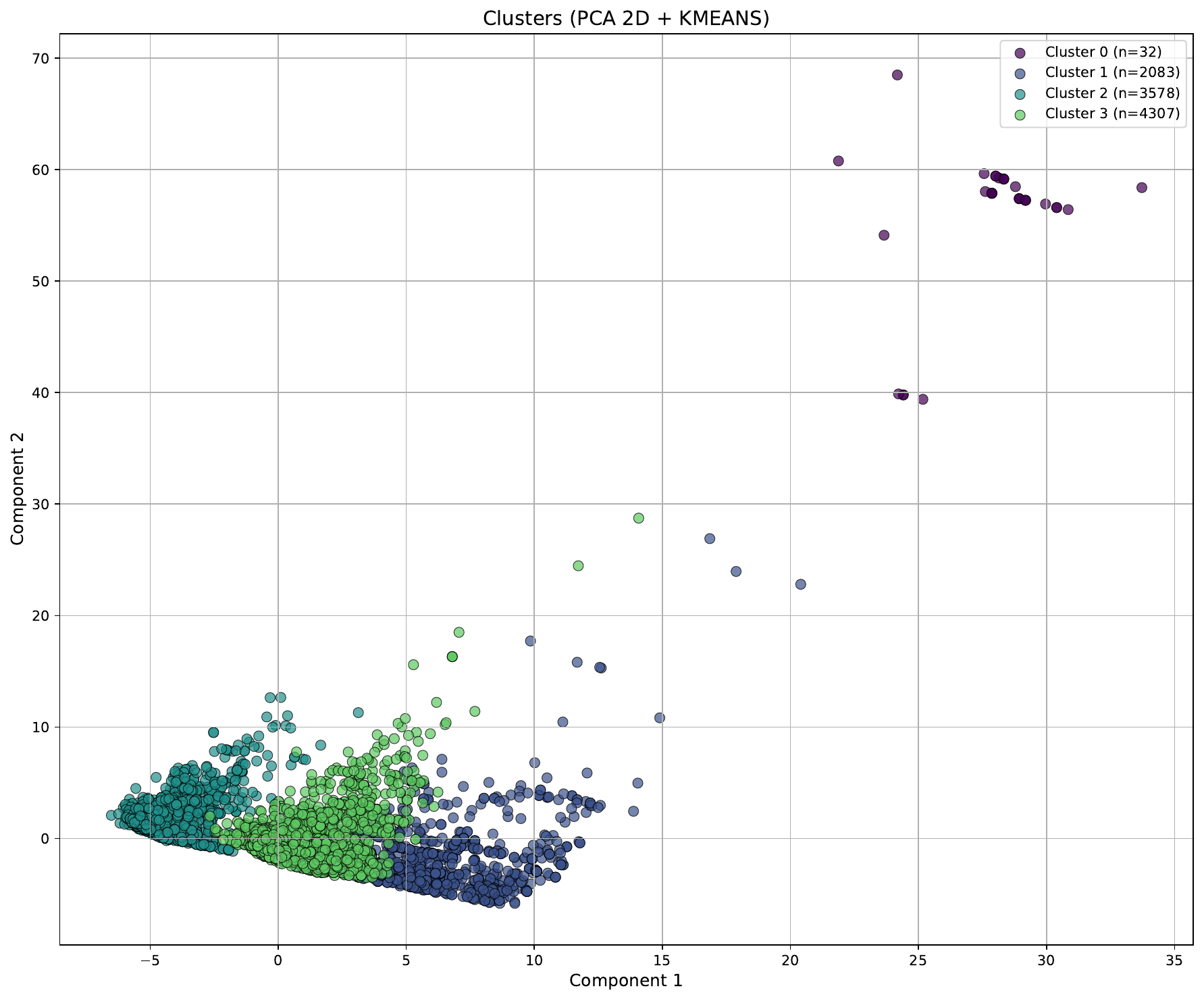}
        \caption{Clustering of malware samples in Kronodroid E dataset.}
        \label{fig:clustering_kronodroid_e}
\end{figure}
\begin{figure}[!htbp]
        \centering
        \includegraphics[width=\linewidth]{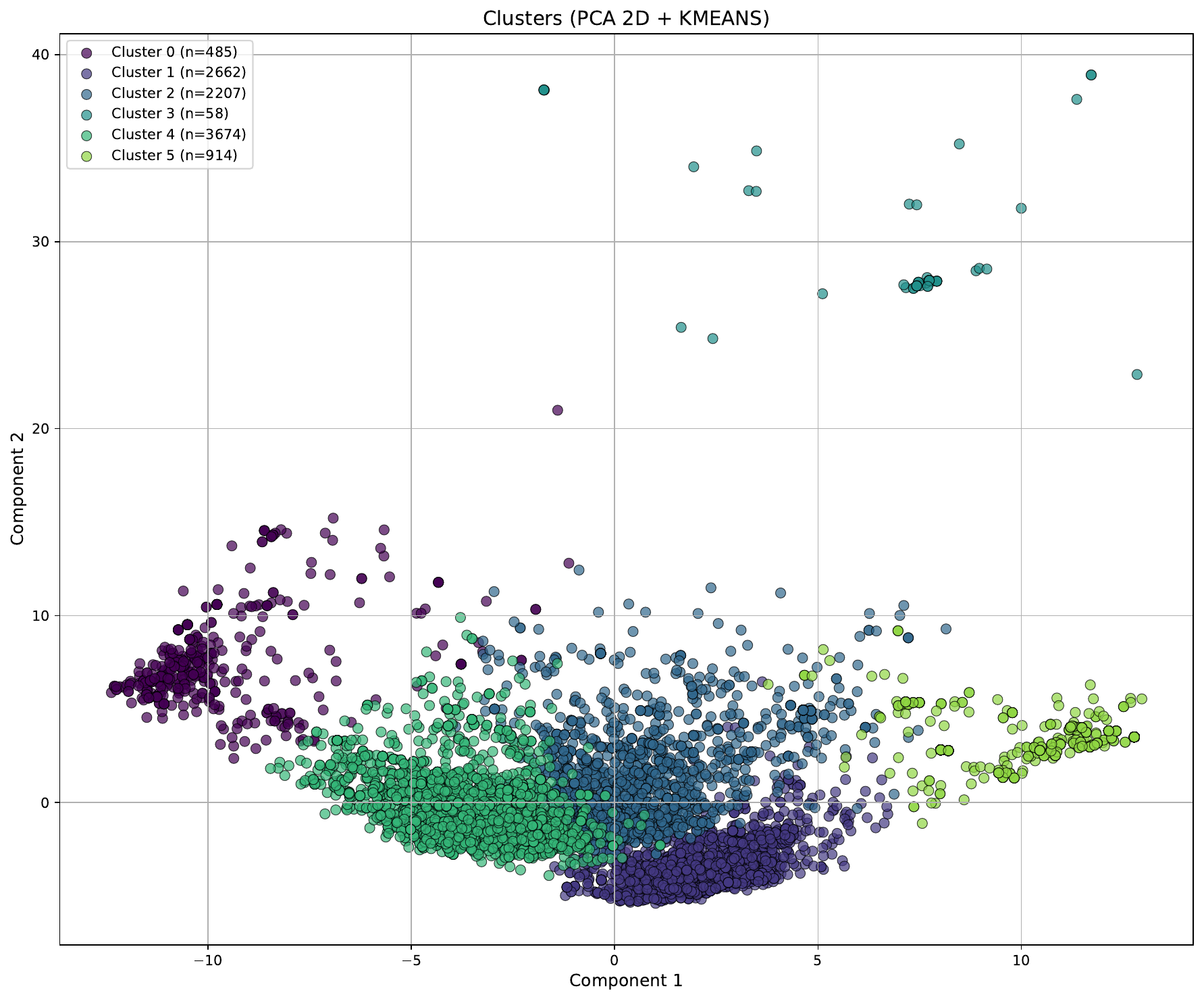}
        \caption{Clustering of malware samples in Kronodroid R dataset.}
        \label{fig:clustering_kronodroid_r}
\end{figure}
\begin{figure}[!!h] 
 
        \centering
        \includegraphics[width=\linewidth]{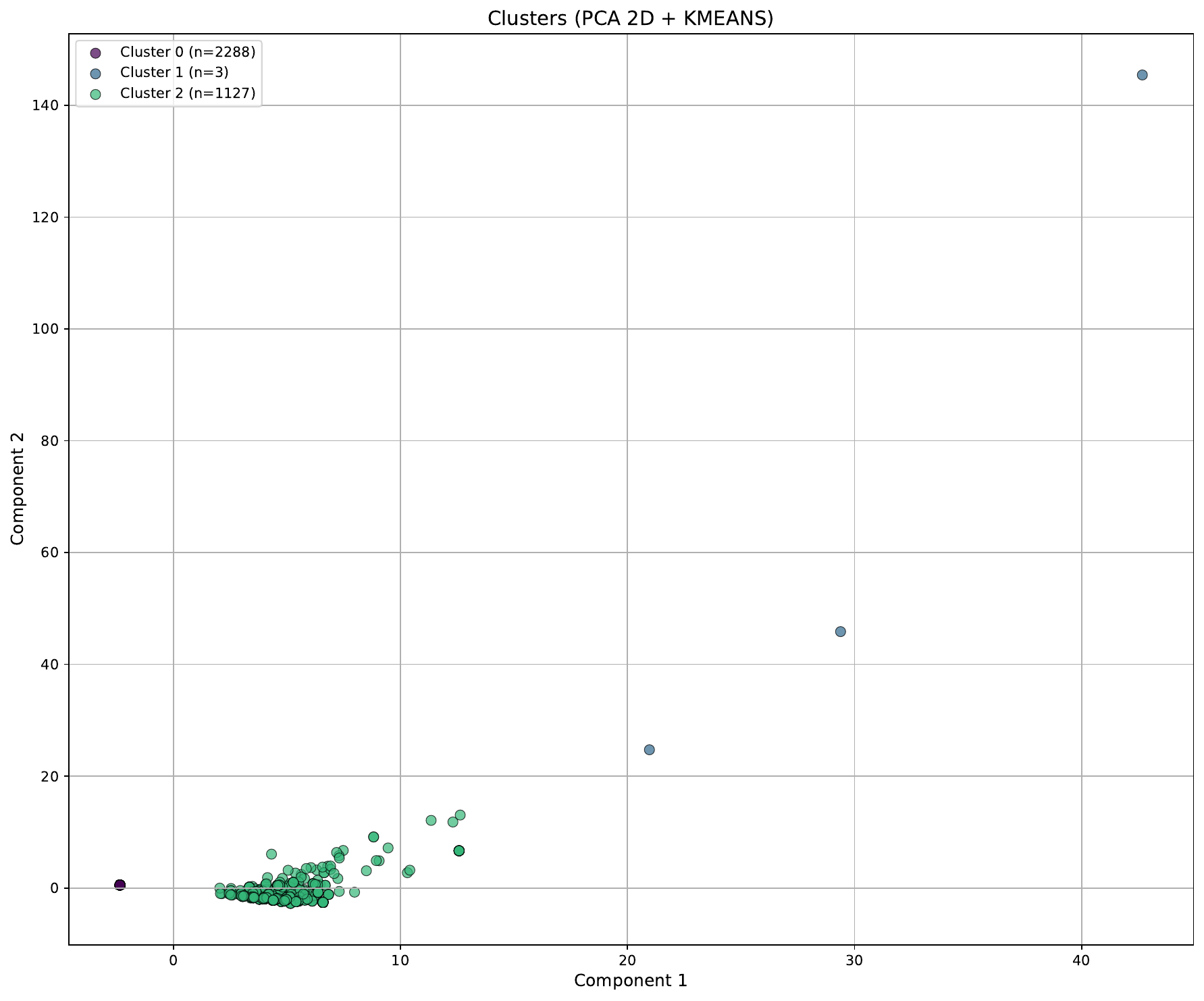}
        \caption{Cluster analysis of malware samples in Adroit dataset.}
        \label{fig:clustering_adroit}
\end{figure} 
\begin{figure}[!h] 
        \centering
        \includegraphics[width=\linewidth]{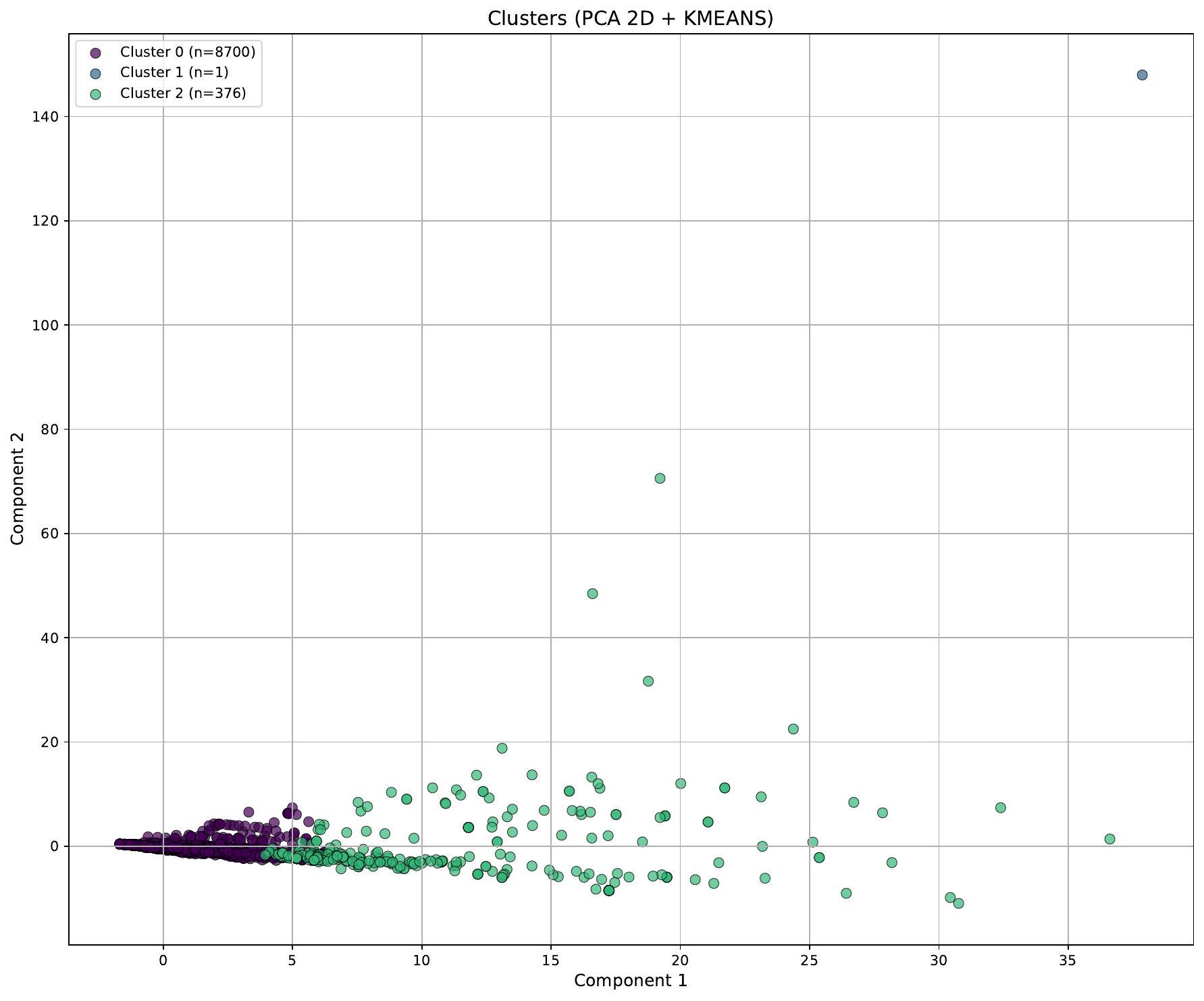}
        \caption{Cluster analysis of malware samples in Android Permission dataset.}
        \label{fig:clustering_android_p}
\end{figure} 

Analyzing the figures, we observed two distinct clustering patterns, similar to those presented in Figure~\ref{fig:clustering}.
In the first pattern, Figures~\ref{fig:clustering_kronodroid_e} and~\ref{fig:clustering_kronodroid_r} exhibit distinct groupings that resemble Figure~\ref{fig:clustering}(a) (Androcrawl). These are characterized by one dominant cluster containing the majority of samples (e.g., Cluster 3, 4,307 in Kronodroid E; Cluster 4, 3,674 in Kronodroid R), alongside several secondary clusters of comparable size. These figures also include small outlier clusters (e.g., Cluster 0, 32 in Kronodroid E; Cluster 3, 58 in Kronodroid R).

The second pattern, observed in the Adroit and Android Permission datasets, is more similar to Figure~\ref{fig:clustering}(b) (Drebin). In these cases, samples and clusters are grouped more tightly together, with a very small number of anomalous samples forming their own distinct clusters (e.g., Cluster 1, 1 sample in Android Permission; Cluster 1, 3 samples in Adroit dataset).

\end{appendices}
\end{document}